\documentclass[preprint]{aastex}
\usepackage{epsfig}
\usepackage{psfig}
\shorttitle{Solar Flare Loop Top Hard X-Ray Emission}
\shortauthors{Petrosian, Donaghy \& McTiernan}

\begin{document}

\title{Looptop Hard X-ray Emission in Solar Flares: Images and Statistics}
\author{Vah\'{e} Petrosian\altaffilmark{1}, Timothy Q. 
Donaghy\altaffilmark{2}}
\affil{Center for Space Science and Astrophysics\\
Stanford University\\
Stanford, CA 94305-4060\\
and \\}
\author{James M. McTiernan}
\affil{Space Sciences Laboratory\\
University of California, Berkeley}

\altaffiltext{1}{Also Departments of Physics, Applied Physics and Astronomy 
Program}
\altaffiltext{2}{Current address Department of Physics, University of Chicago}

\newcommand{\beq}{\begin{equation}}
\newcommand{\eeq}{\end{equation}}
\newcommand{\lsim}{\mbox{$\stackrel{<}{\scriptstyle\sim}$}}
\newcommand{\gsim}{\mbox{$\stackrel{>}{\scriptstyle\sim}$}}
\newcommand{\D}[2]{\makebox{$\displaystyle\frac{\partial{#1}}{\partial{#2}}$}}
\newcommand{\DD}[2]{\makebox{$\displaystyle\frac{\partial^2{#1}}{\partial{#2}^2}
$}}

\begin{abstract} 

The discovery of hard X-ray sources near the top of a flaring loop by the
HXT instrument on board the YOHKOH satellite represents a significant progress 
towards the
understanding of the basic
processes driving solar flares.  In this paper we extend the previous
study of limb flares by Masuda (1994) by including all YOHKOH observations up
through August 1998.  We report that from October 1991 to August 1998, 
YOHKOH observed 20 X-ray bright limb flares (where we use the same selection
criteria as Masuda), of which we have sufficient data to analyze 18 events,
including 8 previously unanalyzed flares.  Of these 18 events, 15 show
detectable impulsive looptop emission. Considering that the finite dynamic 
range (about a decade) of the detection introduces a strong bias
against observing comparatively weak looptop sources, we conclude that looptop 
emission is a common feature of all flares.  We summarize the
observations of the footpoint to looptop flux ratio and the spectral indices.
We present light curves and images of all the important newly analyzed
limb flares. Whenever possible we present results for individual pulses in 
multipeak flares and for different loops for multiloop flares. We then discuss 
the statistics of the fluxes and spectral indices of the looptop and footpoint 
sources taking into account observational selection biases. The importance of 
these observations (and those expected from the scheduled HESSI satellite with 
its superior angular spectral and temporal resolution) in constraining 
acceleration models and parameters is  discussed briefly.

{\it Subject Headings:} 
Sun:flares--Sun:X-rays--acceleration of particles

\end{abstract}

\section{INTRODUCTION}

The discovery of hard X-ray sources located at or above the top of solar flare
loops by the HXT instrument on board the YOHKOH satellite has provided great
insight into the processes that drive solar flares.  The canonical event for 
this
phenomenon is the flare of January 13, 1992, first described by Masuda et al.
(1994), and later analyzed by Alexander \& Metcalf (1997), and is commonly
referred to as the ``Masuda'' flare.  This flare, which is clearly delineated by
a soft X-ray (thermal) loop, shows three compact hard X-ray sources, two located
at the footpoints (FPs for short) and a third above the top of the soft X-ray
loop.  The first systematic study of such sources was undertaken by Masuda
(1994).  During the period of time between the satellite's first scientific
observations (October 1991) up to September 1993, Masuda selected 11 X-ray
bright limb flares observed by YOHKOH that met his selection criteria (see
below), one of these 11 events was interrupted event and was not analyzed
further.  Of the remaining 10 flares, 6 demonstrated a nonthermal looptop (LT
for short) source while another demonstrated what Masuda termed a ``super hot''
thermal LT source.  This indicates that LT hard X-ray emission is fairly common.
This view is strengthened further considering the limited dynamic range of the
HXT instrument and the result we describe in this paper.  Thus, it is reasonable
to conclude that LT sources are present in all flares.

It is generally agreed that the primary energy of solar flares must come from
reconnection of stressed magnetic fields, and as pointed out by Masuda et al.
(1994), the YOHKOH LT observations lend support to theories that place the
location of energy release high up in the corona.  The energy released by
reconnection can be used to heat the ambient plasma and/or to accelerate
electrons and protons to relativistic energies.  The power law hard X-ray
spectra seen in many of the LT sources indicate that nonthermal processes, such
as particle acceleration, are indeed occurring at or near these locations.  The
exact mechanism of the acceleration is a matter of considerable debate.  In
previous works (see Petrosian 1994 and 1996) we have argued that among the three
proposed particle acceleration mechanisms (electric fields, shocks, and plasma
turbulence or waves) the stochastic acceleration of ambient plasma particles by
plasma waves provides the most natural means of explaining the observed spectral
and spatial features of solar flares.

In two recent works (Petrosian \& Donaghy, 1999 and 2000, hereafter {\bf PD}) we
demonstrated how the power-law spectral indices and emission ratios (obtained by
Masuda 1994) can be used to constrain the model parameters.  In this paper, we
expand and extend Masuda's analysis for further investigation of the ubiquity
and nature of the LT source, and to gain a clearer picture of the range of
values of the fluxes and spectral indices of the FP and LT sources.
Furthermore, we investigate the temporal evolutions of the images of several new
flares observed by HXT to determine the relationship between the many spatially
and temporally distinct sources that occur in complex flaring events.  In the
next section we describe our procedure, and in \S 3 we present the light curves
and images of most of the new events, including three events which appear to be
examples of multiple loop flares.  In \S 3 present the statistics of the
relative fluxes and spectral indices of the LT and FP sources.  In \S 5 we
present a summary and our conclusions.

\section{PROCEDURE}

In seeking to expand the sample of flares with looptop emission, we have 
searched 
through ``The 
YOHKOH HXT Image Catalog'' (Sato 
et al., 1998) for appropriate limb flares using Masuda's  two criteria:

1) Heliocentric longitude $>80 \arcdeg$; this ensures maximum angular separation 
between LT and FP sources.

2) Peak count rate in the M2 band $>10$ counts per sec per subcollimator (cts 
s$^{-1}$ SC$^{-1}$, for 
short); this ensures that at least one image can be formed at high energies 
($\sim 33-53$ keV) where thermal contribution is expected to be lower.

In total, we found 20 events from October 1991 through August 1998 
that satisfy these conditions.
Of the events before September 1993 (when Masuda stopped his search), we found 
12 such flares, including 11 events noted by Masuda and one event (that of 
December 18, 1991) which satisfies the search criteria but was apparently 
overlooked by Masuda.
From the period after September 1993 up to August 1998 (when the image catalog 
was finished) we have an additional 8 events.
In the interest of completeness we include all twenty events in this survey (see 
Table \ref{events}).
However, two events (November 2, 1992 noted but not analyzed by Masuda, and May 
10, 1998) have their observations interrupted by spacecraft night, and we are 
unable to obtain any useful information from them.
An interesting aside (which we shall return to in \S 4), is that of the 9 new 
events, 3 appear to be examples of interacting loop structures with multiple 
LT and FP sources, of the type analyzed by Aschwanden et al. 
(1999).
It is surprising that none of the 11 Masuda events are in this category.
It should also be noted that three of the flares observed before September 1993 
show no detectable LT emission, while all of the flares since that date 
(with the exception of one ambiguous source) show LT emission.

For the ten early events, Masuda quotes LT and FP spectral indices 
for the events when those sources are detected, although he does not supply the 
time intervals or spatial regions over which the indices are taken.
Masuda obtains two spectral indices by fitting  a power law  to the L ($\sim 
13-23$ keV) and M1 ($\sim 23-33$ keV)
band fluxes, and the M1 and M2 band fluxes.
We carry out spectral fitting both to a simple power law  as well as a broken 
power law  using all four channel fluxes (including the H band with energy range 
$\sim 53-93$ keV), whenever significant fluxes in all four 
bands are detected. These results are summarized in  Tables \ref{2} and 
\ref{3}.

For two events (January 13, 1992 and October 4, 1992), Masuda performs a 
temporal and spatial analysis, providing light curves for emission from the 
LT and FP regions.
He defines a box around each source, with size and center position determined 
from the best available image, and then obtains light curves by measuring the 
total flux of photons coming from within the box as a function of time.
We undertake this temporal analysis for all the events in our list for which the 
necessary data are available.
The data reduction is done with the standard YOHKOH HXT software package, 
which uses the maximum entropy method (MEM for short); in particular we use \sf 
hxt\_multimg, lcur\_image, \rm and \sf hxt\_boxfsp \rm routines.

\begin{table}[ht] \centering \scriptsize
\begin{tabular}{|c|c|c|c|c|r|c|}\hline\hline
Event \# & Date & Peak Time & Disk & GOES & L band  & Loop Top \\
& & hhmmss & Position & Class & Peak Count & Location  		\\
\hline
1 & 91/12/02 & 045527 & N18E84 & M3.6 & 61    & inside*        \\	
2 & 91/12/15 & 024411 & ------ & M1.2 & 30    & NO SXT 		\\
3 & 91/12/18 & 102740 & S14E84 & M3.5 & 379   & inside(New)	\\
4 & 92/01/13 & 172937 & S16W84 & M2.0 & 51    & above*		\\
5 & 92/02/06 & 032511 & N06W84 & M7.6 & 807   & inside(s-hot)   \\
6 & 92/02/17 & 154209 & N16W80 & M1.9 & 33    & inside	        \\
7 & 92/04/01 & 101407 &  ------ & M2.3 & 50   & no LT source*	\\
8 & 92/10/04 & 222107 & S05W90 & M2.4 & 35    & above	        \\
9 & 92/11/02 & 030002 & S24W84 & X9.0 & 11608 & Int.	        \\
10 & 92/11/05 & 061959 & S18W84 & M2.0 & 68   & none		\\
11 & 92/11/23 & 202602 & S08W84 & M4.4 & 94   & NO SXT		\\
12 & 93/02/17 & 103630 & S07W87 & M5.8 & 229  & above		\\
\hline
13 & 93/09/27 & 120839 & N08E84 & M1.8 & 70   & above		\\
14 & 93/11/30 & 060337 & S20E84 & C9.2 & 57   & C		\\
15 & 98/04/23 & 054454 & S20E84 & X1.2 & 1496 & single*		\\
16 & 98/05/08 & 015847 & S16W84 & M3.1 & 116  & C		\\
17 & 98/05/09 & 032641 & S16W84 & M7.7 & 860  & NO SXT		\\
18 & 98/05/10 & 131849 & S28E84 & M3.9 & 138  & Int.	        \\
19 & 98/08/18 & 082141 & N34E84 & X2.8 & 3499 & inside?		\\
20 & 98/08/18 & 221534 & N30E84 & ---- & 12397& C		\\ 	
\hline 
\hline
\end{tabular}
\caption{List of events from October 1991 to August 1998 that satisfy 
Masuda's two criteria.
Dates, UT times, position on the solar disk, flare classifications and the Peak 
Counts in the L band (columns 2 to 6) are taken from Sato et al. (1998). Column 
7 describes the location of the LT HXT source relative to the SXT loop and some 
other characteristics of the images. The horizontal line divides the Masuda 
(pre September 1993) flares and our post September 1993 flares.
Flares identified by a * may be partially occulted by the solar limb.
NO SXT implies absence of SXT image of the loop.
Flare \# 3 labeled  (New) was missed by Masuda. Flare \# 4 was classified by 
Masuda as a super-hot thermal source. 
Flare \# 15 (identified as single*) shows presence of only one resolved source 
which may be the LT source of an occulted loop.
C=complex multiple loop event.
Int.= Interrupted observation, not analyzed.}
\label{events}
\end{table}

For several events, we have performed image reconstruction using both the 
standard MEM  routine as well as the ``pixon'' technique 
used by Alexander \& Metcalf (1997).
As described in that paper, the ``pixon'' routine achieves significantly better 
photometry of faint sources and better rejection of spurious sources than the 
standard MEM routine, albeit with a significantly longer computational time.
These features are useful to our analysis since we are often attempting to 
resolve a faint LT source in the presence of much stronger FP 
sources.
In general, the ``pixon'' and MEM analyses yielded  similar light curves.

Major sources of systematic uncertainty in our analysis arise from the 
difficulty in defining exactly what constitutes a LT or a FP 
source.
In many cases, one or more of these sources will be quite strong and well 
localized at one epoch during a flare but they may disappear, change position 
or combine with another source as the flare evolves.
Therefore, the procedure of integrating the fluxes using a static box to obtain 
a light curve for the duration of the flare can lead to erroneous conclusions 
because of the possibility of contamination of one source by neighboring 
sources.
Further errors arise from the fact that in many flares nonthermal emission, 
which tends to be more 
impulsive  and and peak earlier is superimposed upon the more gradual thermal 
emission which peaks later. Therefore for the study of the 
impulsive component (as a means of studying the acceleration mechanism) we limit 
our analysis of the 
flux ratios or spectral indices to the early stages of the flares to minimize 
the influence of the thermal component.
We also use the M1 channel (rather than the L channel)  where the thermal 
effects are expected to be smaller. Clearly it would be safer to use channels M2 
and H, but these channels have weaker emissions.

\section{CASE STUDIES OF IMAGES AND LIGHT CURVES}

In this section we present light curves and a brief analysis for all of the 
newly analyzed events.
Unless otherwise noted, the image and light curves were taken in the M1-band and 
using the standard MEM procedure.

\subsection{Single Loop Flares}

Of the 18 total events, 11 appear to by morphologically similar to the canonical 
``Masuda event'', that is, a single flaring loop anchored by twin FPs and 
exhibiting a distinct LT source.
Of course, within this type we observe a good deal of variation. We present a 
few case studies.

\begin{figure}[htb]
\epsfig{file=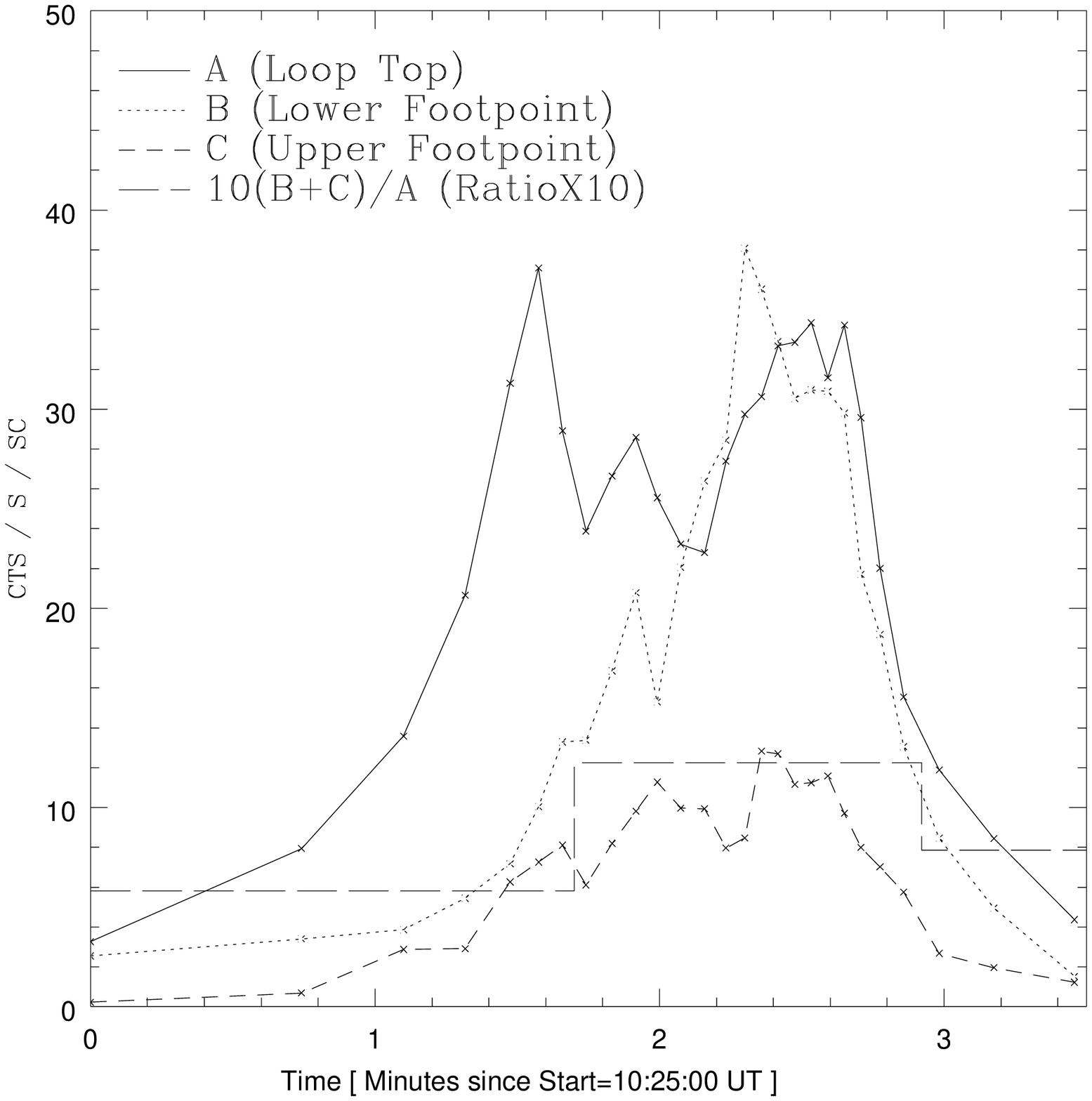,height=3in}
\psfig{file=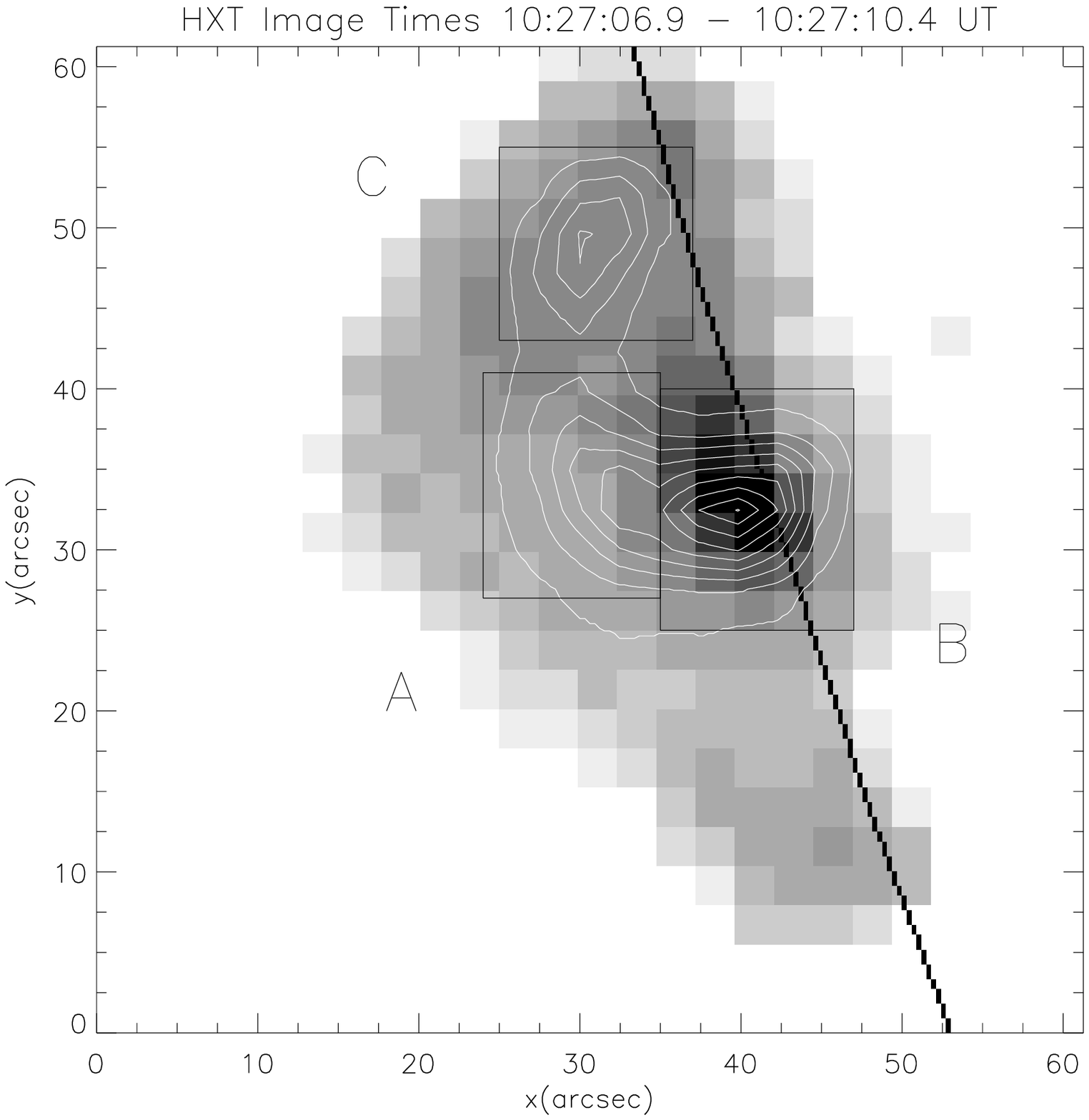,width=3.0in,height=4.0in,angle=90}
\caption{Images ({\bf right panel}) and  light curves ({\bf left panel})  for 
the December 18, 1991 flare. The
contours and the gray scale show the HXT (channel M2; 33-53 keV) and SXT images 
of the loop, 
respectively, for the specified time. The diagonal line shows the location of 
the solar limb. The brightest contour in the HXT image is at $B_{max}=8.1$ 
counts/pixel with 2.5 sq arc second size pixels, and the contour separations are 
$\Delta B=0.73$ counts/pixel. The light curves of the  the LT and FP sources 
refer to the counts integrated over 
regions shown on  right panel. The 
dashed histogram shows the average of the ratio of counts of two FPs to the  LT
sources, ${\cal R}=FPs/LT$ (multiplied by  10) for three time intervals.}
\label{911218}
\end{figure}

{\it December 18, 1991} -- This event occurred during the early phases of the 
YOHKOH mission, but for reasons unknown to us it was not selected or analyzed by 
Masuda.
The hard X-ray images initially show one bright source (at around 10:25:00), 
which could be either a LT or a FP source, coaligned with the 
bright region of the soft X-ray flaring loop.
Shortly after the start of the flare, the single source splits into two, and 
sometimes three, unresolved sources coaligned with the flaring 
loop. The loop joining sources A, B and C, in the right panel of Figure 1, is 
most likely an asymmetric loop. This can  give rise to the observed differences 
between the intensities and heights above the solar limb of the two FP sources B 
and C. The left panel of
Figure \ref{911218}  shows the light curves for each of these three 
regions. The dominance of the  source A during the initial peak 
of the flare could be due to efficient trapping of the accelerated electrons at 
the acceleration site (assumed to be at the  
top of the loop) by a high density of plasma turbulence. Therefore, we 
assume it to be a LT source for this analysis. Note that for this flare the 
ratio of the sum of the counts of the two FPs to that of the LT source, ${\cal 
R}=FPs/LT$, is about one, changing from 0.6 at the beginning to $\sim 1.2$ 
during the main pulse. In our statistical analyses described in the next section 
we will often use the counts $FPs$ and $LT$ of the main pulse, and in few cases 
we use the counts of well defined pulses separately. We will rarely use the 
counts or the ratios obtained during the decaying phase where thermal 
contamination may become large. 
The ``pixon'' reconstruction of the same event shows cleaner images, but 
the same light curves.

\begin{figure}[htb]
\epsfig{file=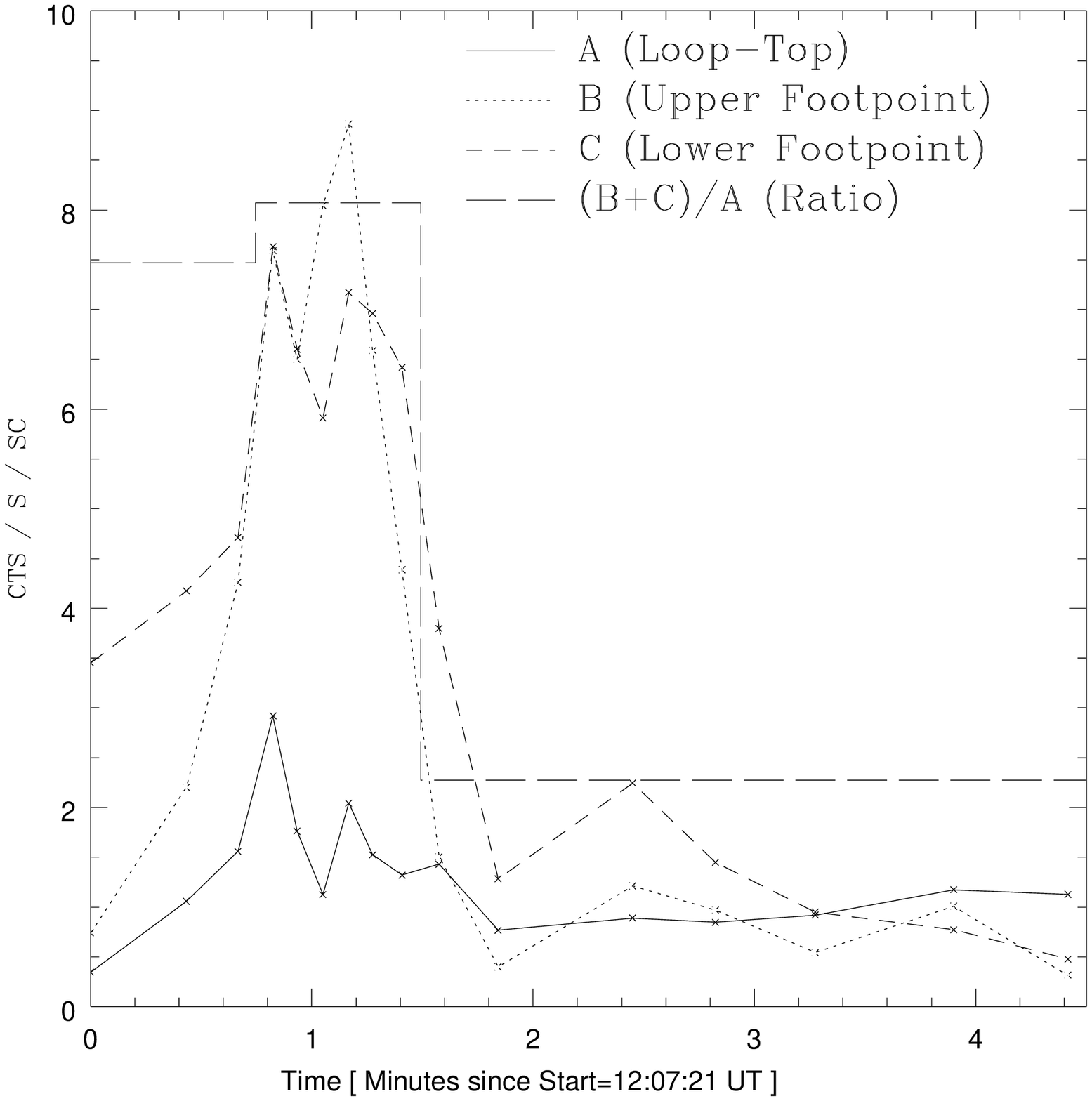,height=3in}
\psfig{file=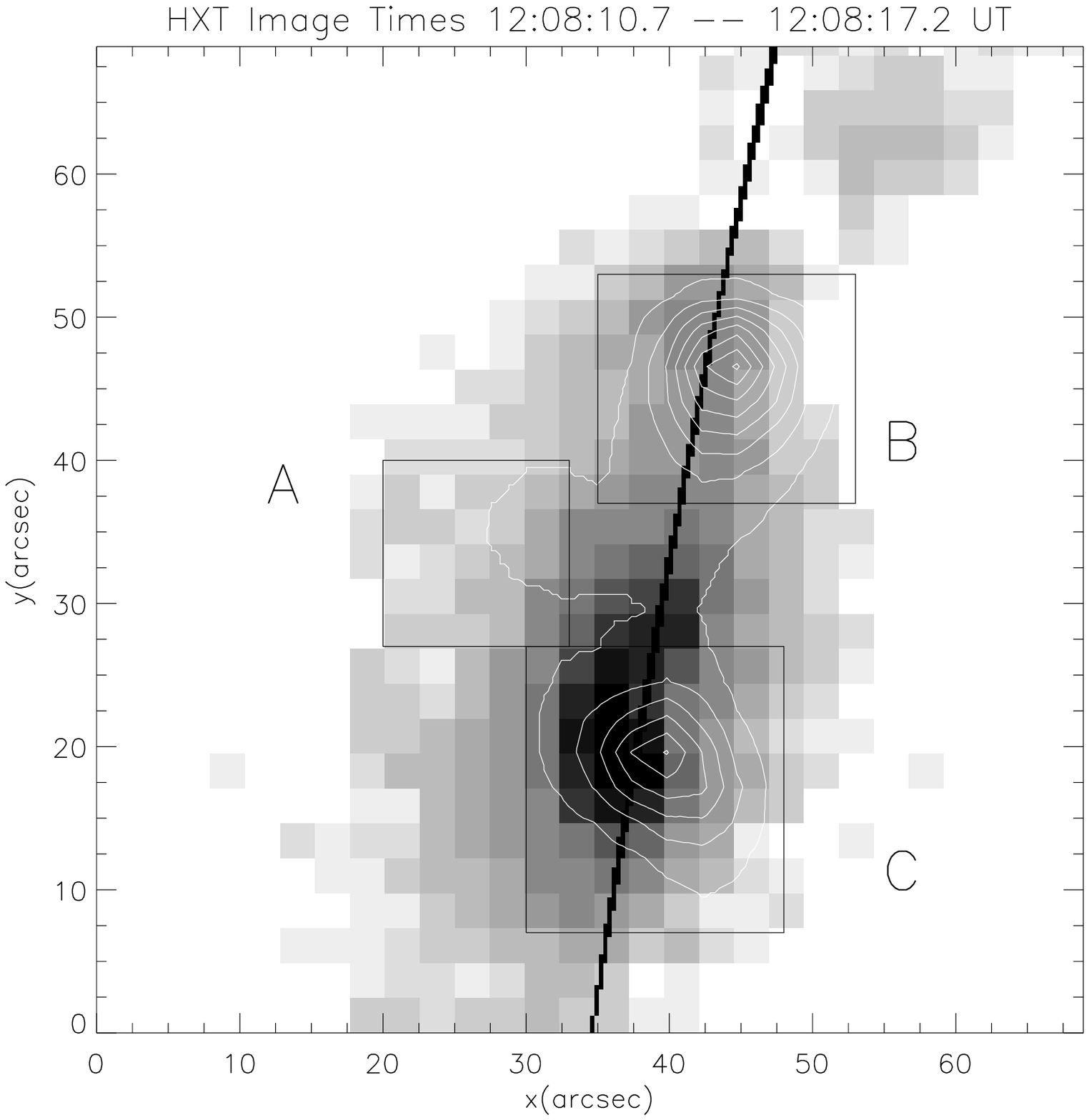,width=3.0in,height=4.0in,angle=90}
\caption{Same as Figure \ref{911218} for the September 27, 1993 flare, with 
$B_{max}=4.1$  and  $\Delta B=0.41$ counts/pixel.}
\label{930927}
\end{figure}

{\it September, 27, 1993} -- The SXT images for this event show a compact 
flaring loop on the limb and the HXT images primarily show two bright FPs 
associated with this loop.
The LT source is weaker (the count ratio ${\cal R} \sim 8$) and appears only 
intermittently  high above the 
flaring loop 
in the corona. However, as shown in
Figure \ref{930927} the LT and FP sources have similar 
pulse structures.

\begin{figure}[htb]
\epsfig{file=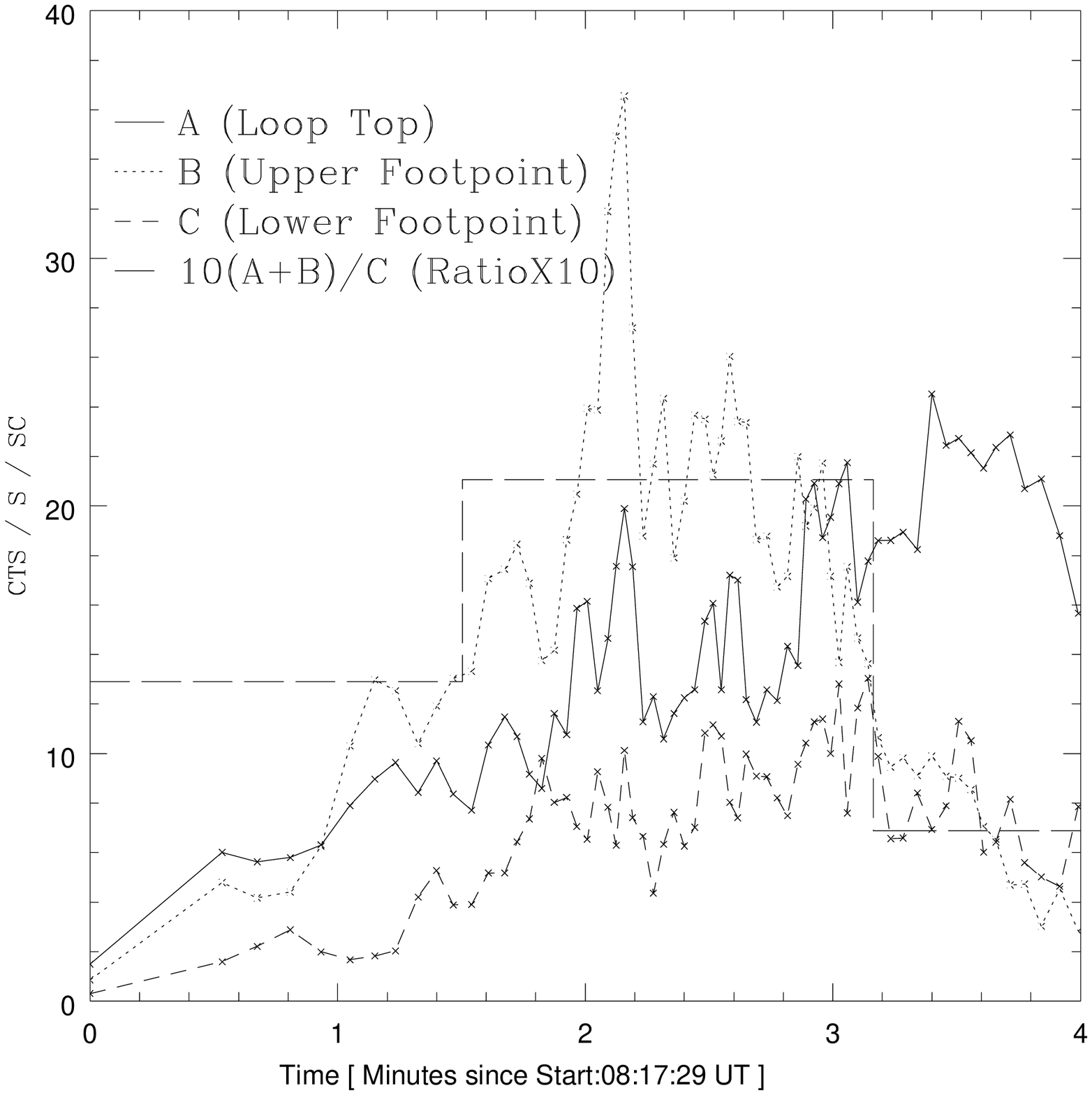,height=3in}
\psfig{file=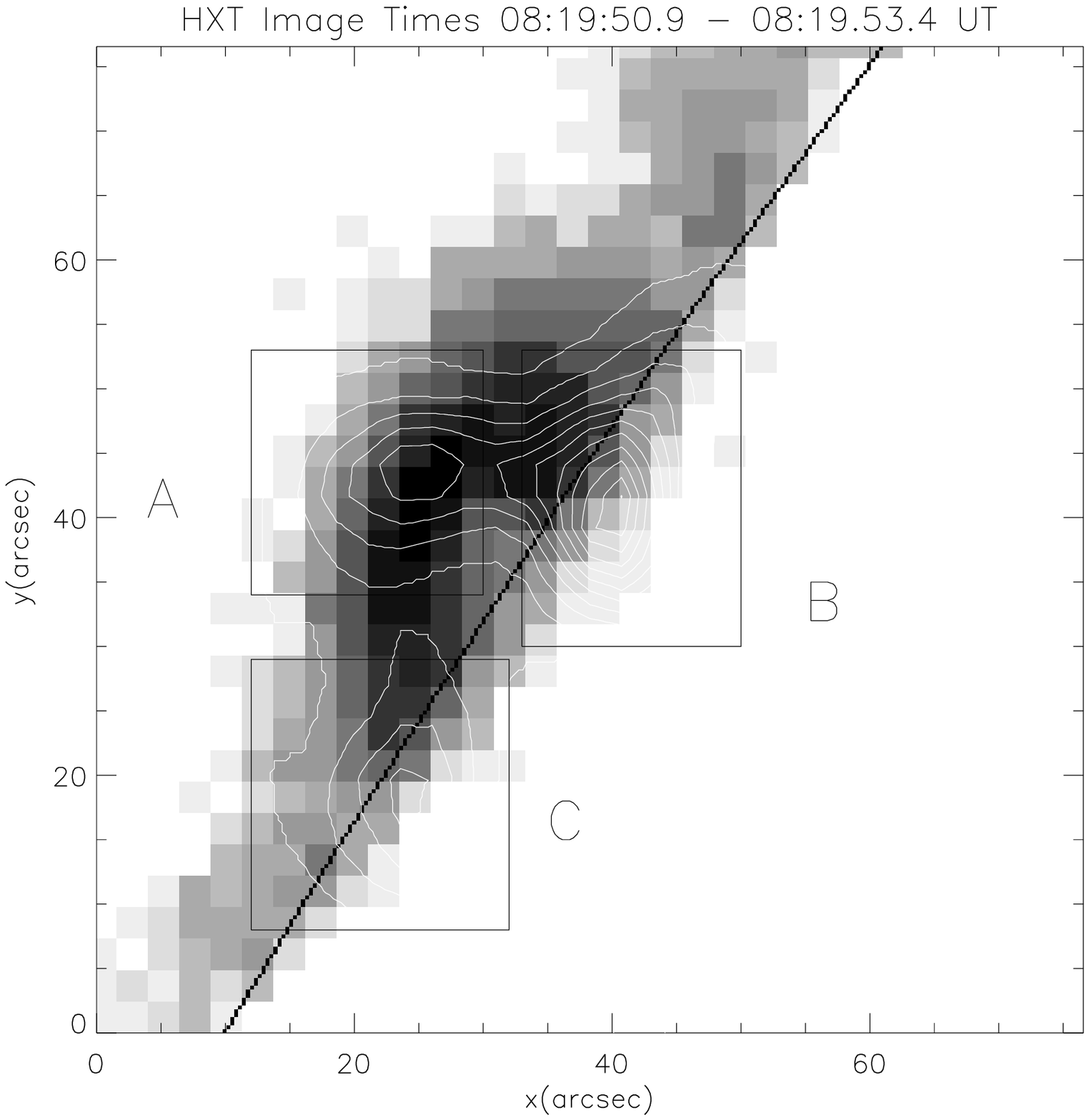,width=3.0in,height=4.0in,angle=90}
\caption{Same as Figure \ref{911218} for the first August 18, 1998 flare, with 
$B_{max}=1.9$  and  $\Delta B=0.15$ counts/pixel.}
\label{a980818}
\end{figure}

{\it August 18, 1998, 08:21UT} -- This very bright X-class flare is the first of 
two bright flares from this active region on this day.
The SXT images show a bright loop structure that is coaligned with two FP 
HXT sources and one bright LT HXT source, Figure \ref{a980818}. 
These three sources persist throughout the impulsive phase of the flare and 
share common impulsive peaks.
In the late phase (after 08:20:15), the LT comes to dominate and the HXT 
emission becomes more localized.
In contrast to the preceding flare, the LT source here is extremely bright and 
long lived.
 
Taken together with the ``Masuda flare'' of January 13, 1992 and the other 
single loop flares analyzed by Masuda, we note that the LT can manifest 
itself in many different forms.
Some flares only show LT emission very briefly or very faintly.
In larger flares, we observe LT sources that are clearly spatially 
separated from the FP sources, yet have similar impulsive time structure.

\subsection{Multiple Loop Flares}

Three of the post 1993 flares we have analyzed show HXT (and SXT) images with 
complex 
morphologies that have strong fluxes coming from several distinct regions which 
brighten and dim at different times. 
These flares cannot be modeled by a single loop and provide evidence for 
existence of multiple loop structures. It should be noted that none of the pre 
1993 flares showed such complex structures.

\begin{figure}[htb] 
\epsfig{file=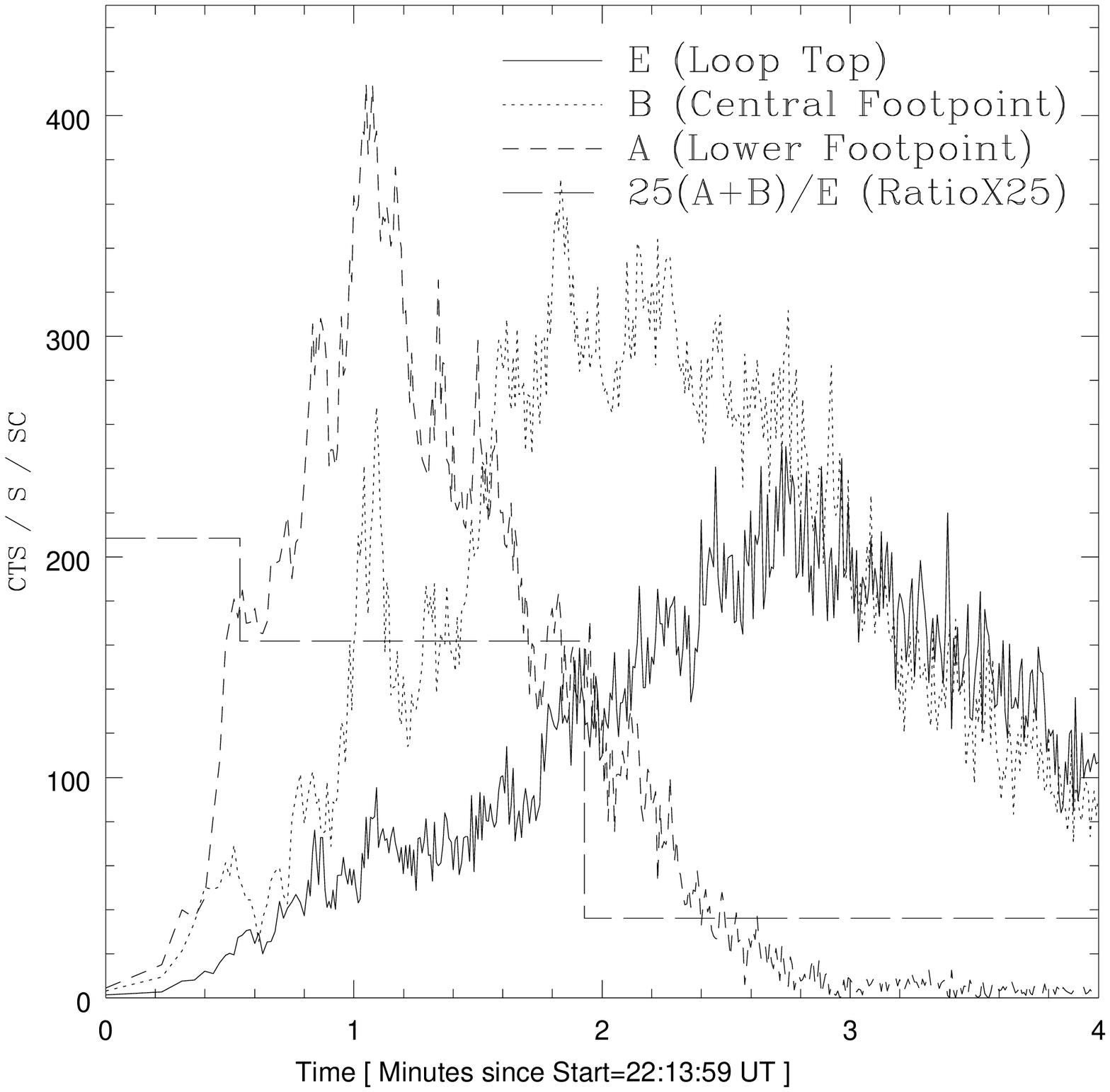,width=3in,height=3in}
\epsfig{file=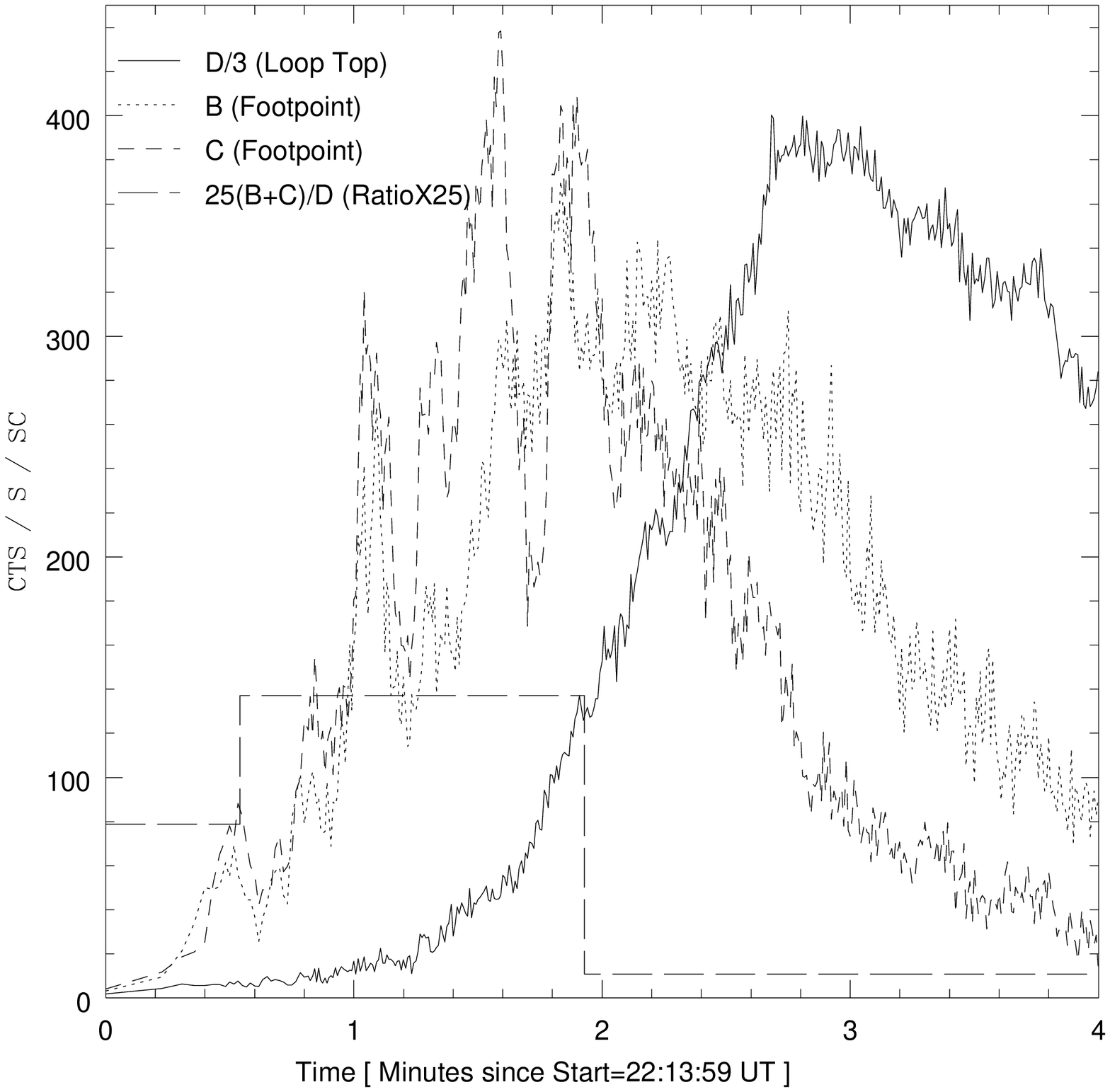,width=3in,height=3in}
\psfig{file=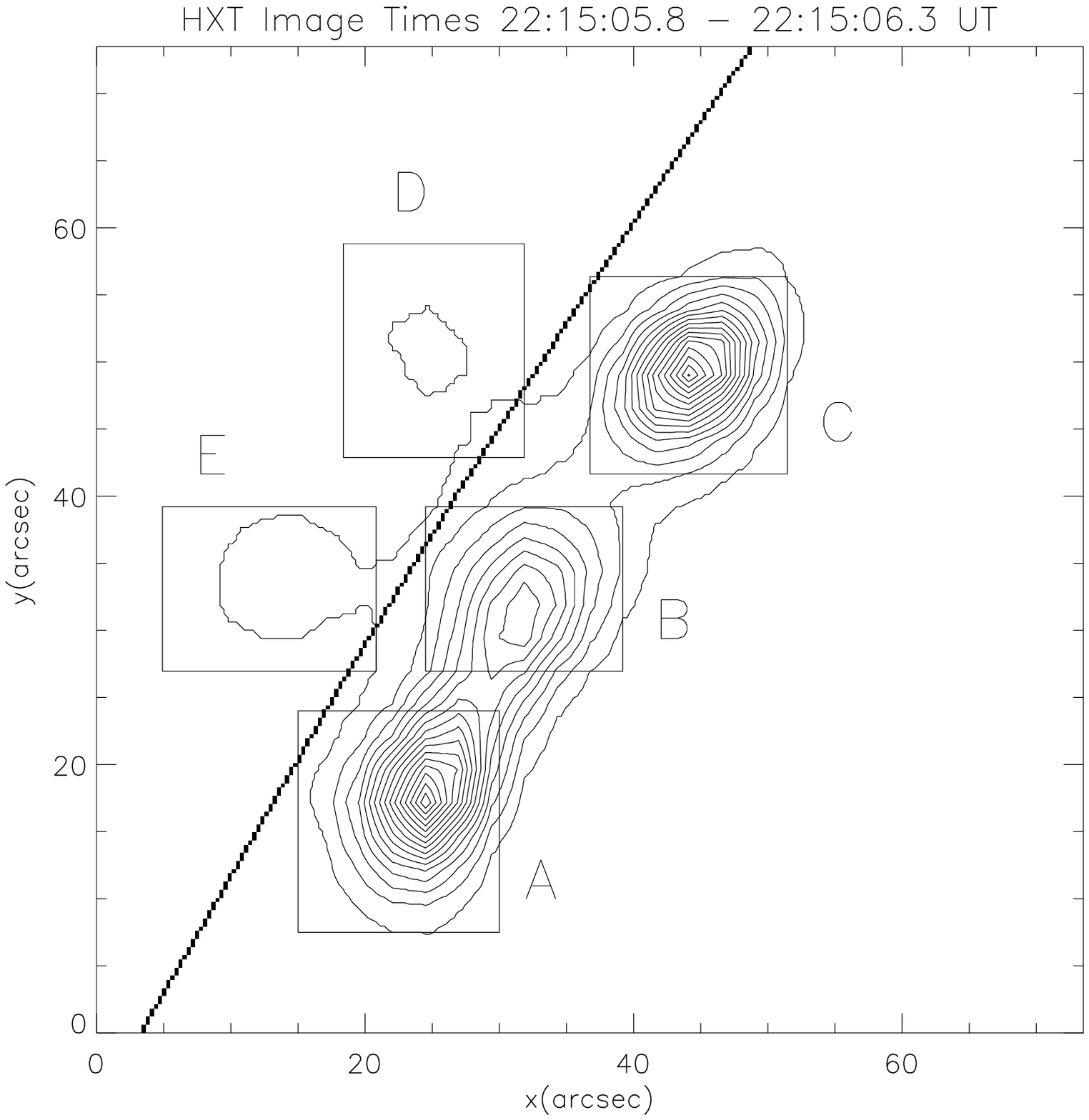,width=2.9in,angle=90}\hspace{-1.0cm}
\psfig{file=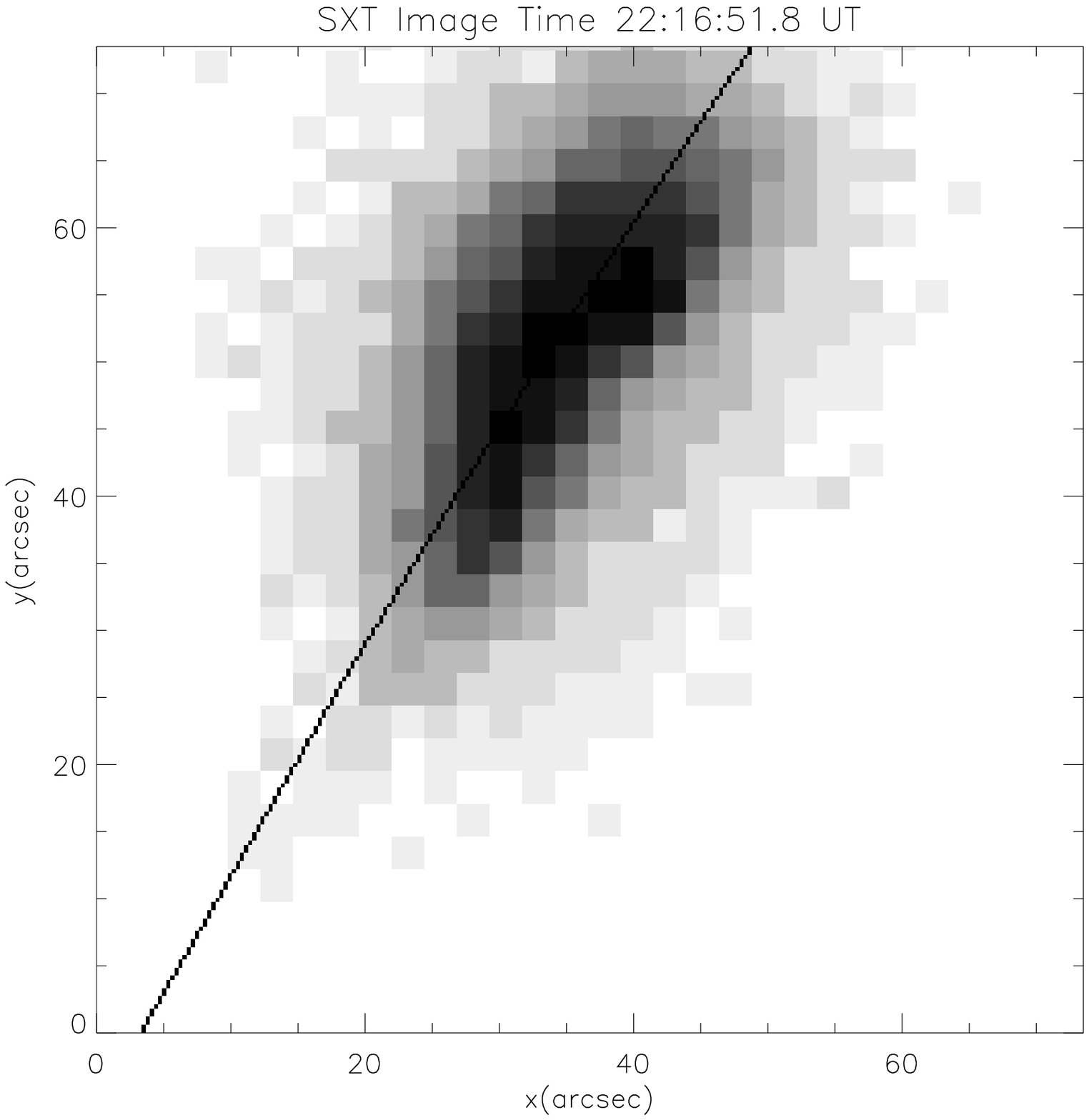,width=2.9in,height=3.8in,angle=90} 
\caption{Same as Figure \ref{911218} for the second August 18, 1998 
flare. The upper left and right panel light curves represent the
southern (AEB)  and  the  northern (BDE) loops, respectively. Note that
for the LT source D we plot counts  divided  by 3.  In the HXT image
(lower left panel) $B_{max}=14.8$ and $\Delta B=0.82$  counts/pixel.
The SXT  image is shown separately on the lower right  panel to avoid
confusion. (Note that the SXT image is taken nearly two minutes  after
than the HXT image. This can explain the misalignments between the two 
images.) The digonal lines in the images show the location of the
limb.}
\label{b980818} 
\end{figure}

\emph{August 18, 1998, 22:15 UT} - This event (see Figure \ref{b980818}) is the 
second flare occurring on this day in this active region, and it is one of the 
most energetic flaring events observed since YOHKOH's launch.
This X-class flare appears in the hard X-rays as two flaring loops sharing a 
common central FP, but is too bright for most of the flare duration to 
yield useful images in soft X-rays.
The early SXT images have many overexposed pixels and the later images show only 
one flaring loop associated with the upper two FPs.
The upper loop is associated with two bright FPs (B and C), and a loop 
top source (D),
which shows impulsive behavior superimposed upon an extremely hot thermal 
component (seen even in the M2 band).
A third FP source (A) lies further south from the soft X-ray flaring loop 
and 
is associated with a comparatively fainter LT source (E) situated between 
it 
and the middle FP.
The light curves give evidence for multiple acceleration sites.
For example, while all three FPs participate in the first large peak 
(just after 22:15:05), only the upper sources show strong emission at the later 
peaks (22:15:30 and 22:15:50), while the lower sources fade away.

\begin{figure}[htb]
\epsfig{file=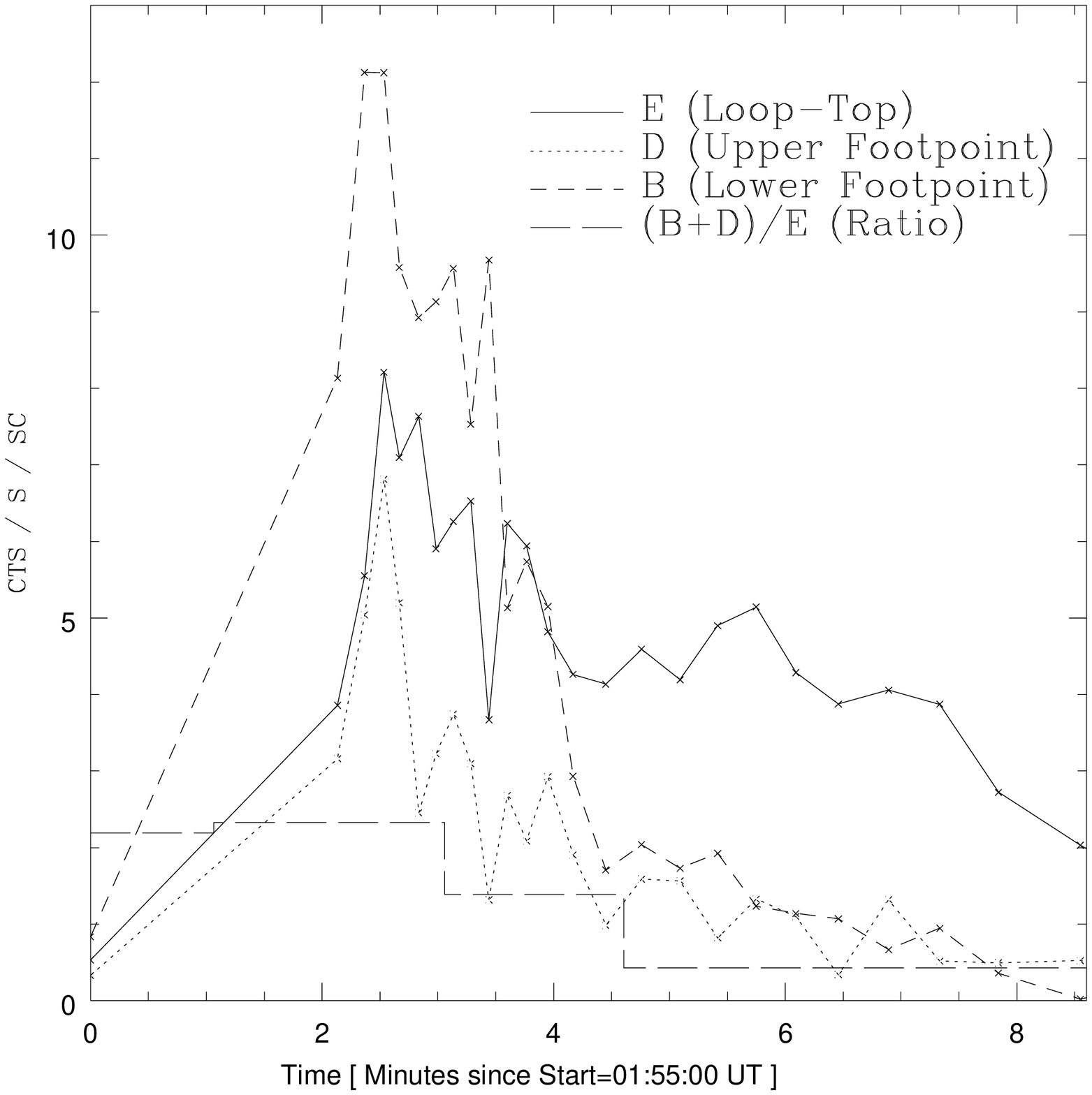,height=3in}
\epsfig{file=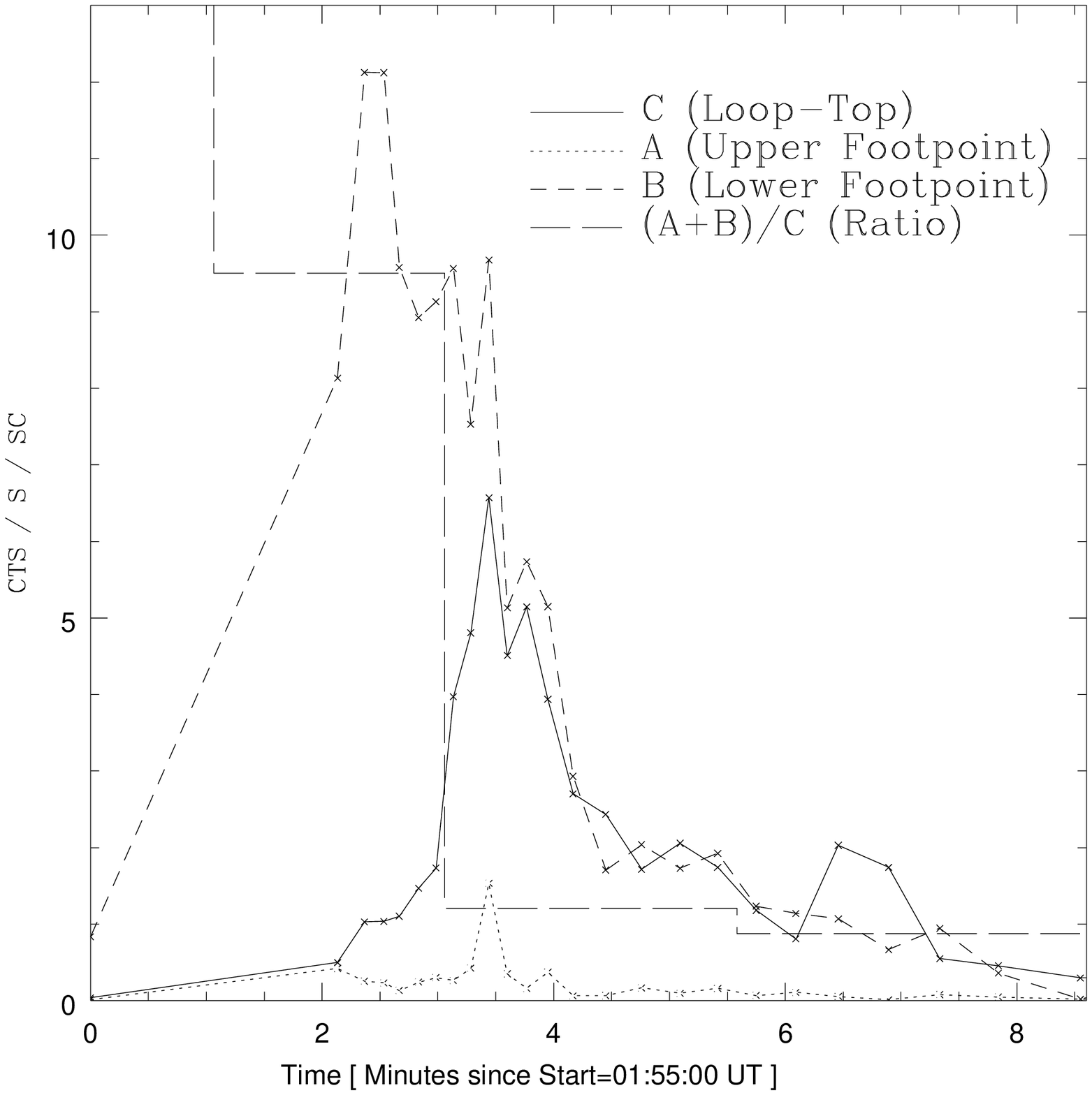,height=3in}
\psfig{file=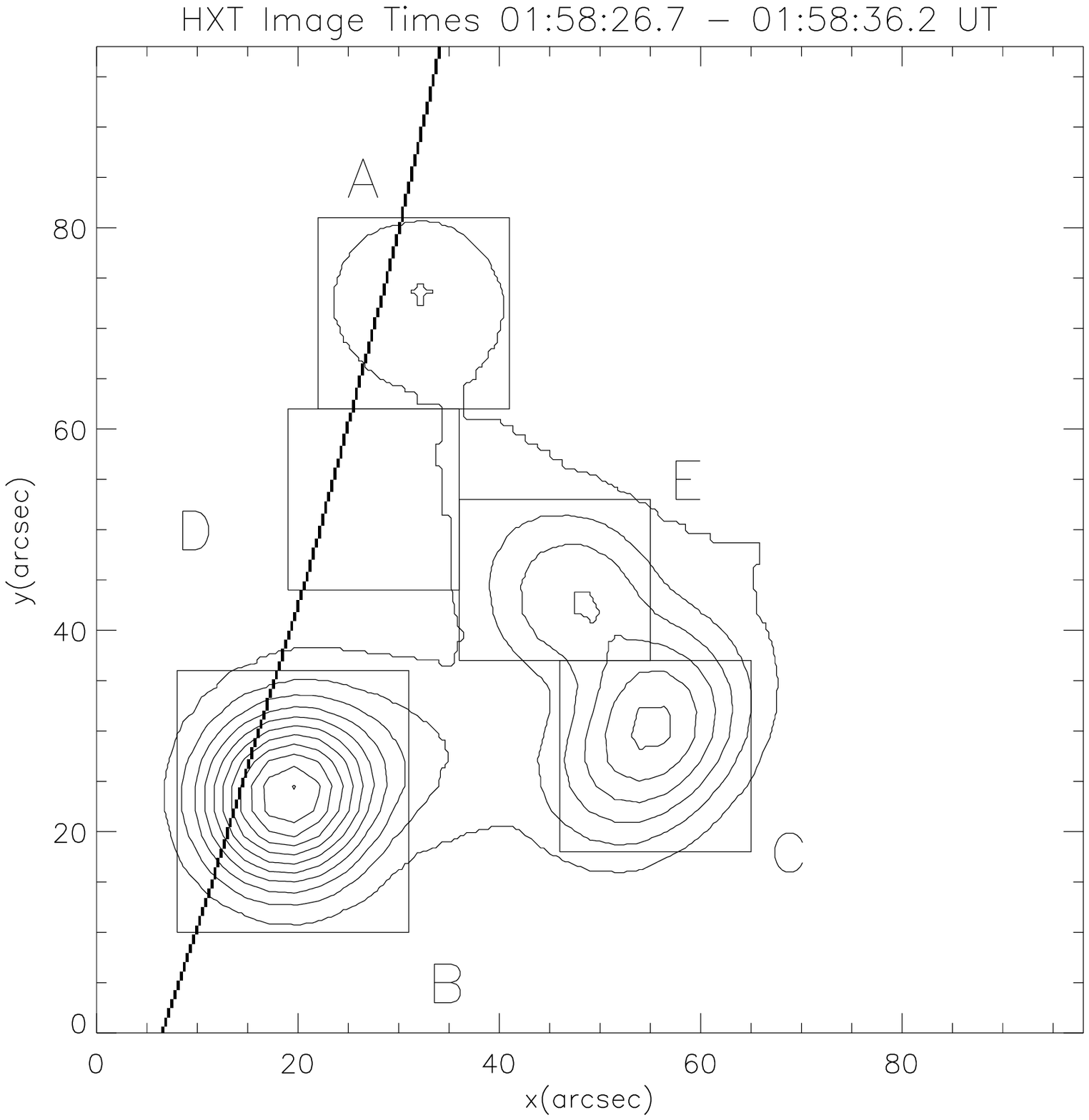,width=2.9in,angle=90}\hspace{-1.0cm}
\psfig{file=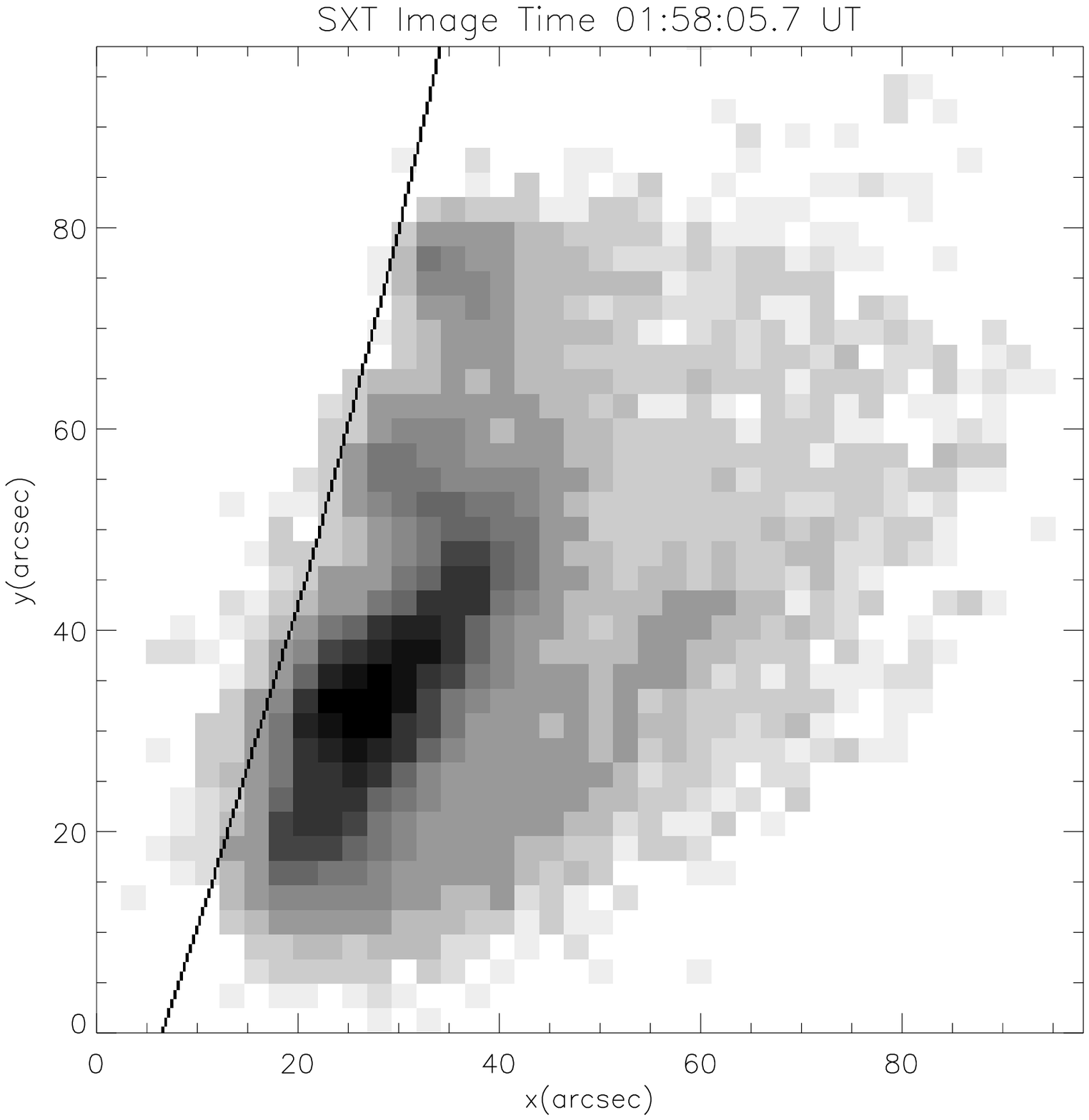,width=2.9in,height=3.8in,angle=90}
\caption{Same as Figure \ref{b980818} for the May 8, 1998 flare.
The upper left and right panel light curves represent the inner (DEB) and the 
outer (ACB)
loops, respectively. In the HXT image $B_{max}=3.0$ and   $\Delta B=0.23$ 
counts/pixel. These images were reconstructed using the ``pixon'' 
method.}
\label{980508}
\end{figure}

\emph{May 8, 1998} - Figure \ref{980508} shows hard X-ray image (at 
01:57:40) and the corresponding light curves for this flare.
The light curves and images presented here are from  the 
``pixon'' reconstruction method, which can identify faint sources better than 
the MEM method.
Evidence from the soft X-ray images points to the existence of two flaring loops 
which share the lower FP (denoted B in the figure) and whose LT 
sources (E and C) overlap.
The smaller loop is brighter and is associated with the FPs B and D and 
the LT source, E.
The fainter, outer loop is associated with B and the considerably fainter 
FP source, A, as well as the LT source C.
While two upper FP sources (D and A) are indeed fainter, they do 
demonstrate the same impulsive variation as their brighter LT and 
FP relatives.
Looking at the light curves, we can see that the inner loop is strongly 
associated with the first impulsive peak at 01:57:30.
The outer loop peaks later (at 01:58:30), at a time which coincides with 
a low point in the light curves of the inner loop sources. It is difficult to 
demonstrate the presence of a causal connection, if any, between these loops.
The time profile of source B is approximately a superposition of the time 
profile of the two 
FPs  A and C that are separated
spatially and temporally.

\emph{November 30, 1993} - This event appears as a double loop structure in soft 
X-rays with apparent multiple LT and FP sources appearing in the 
hard X-ray images.

\begin{figure}[htb]
\epsfig{file=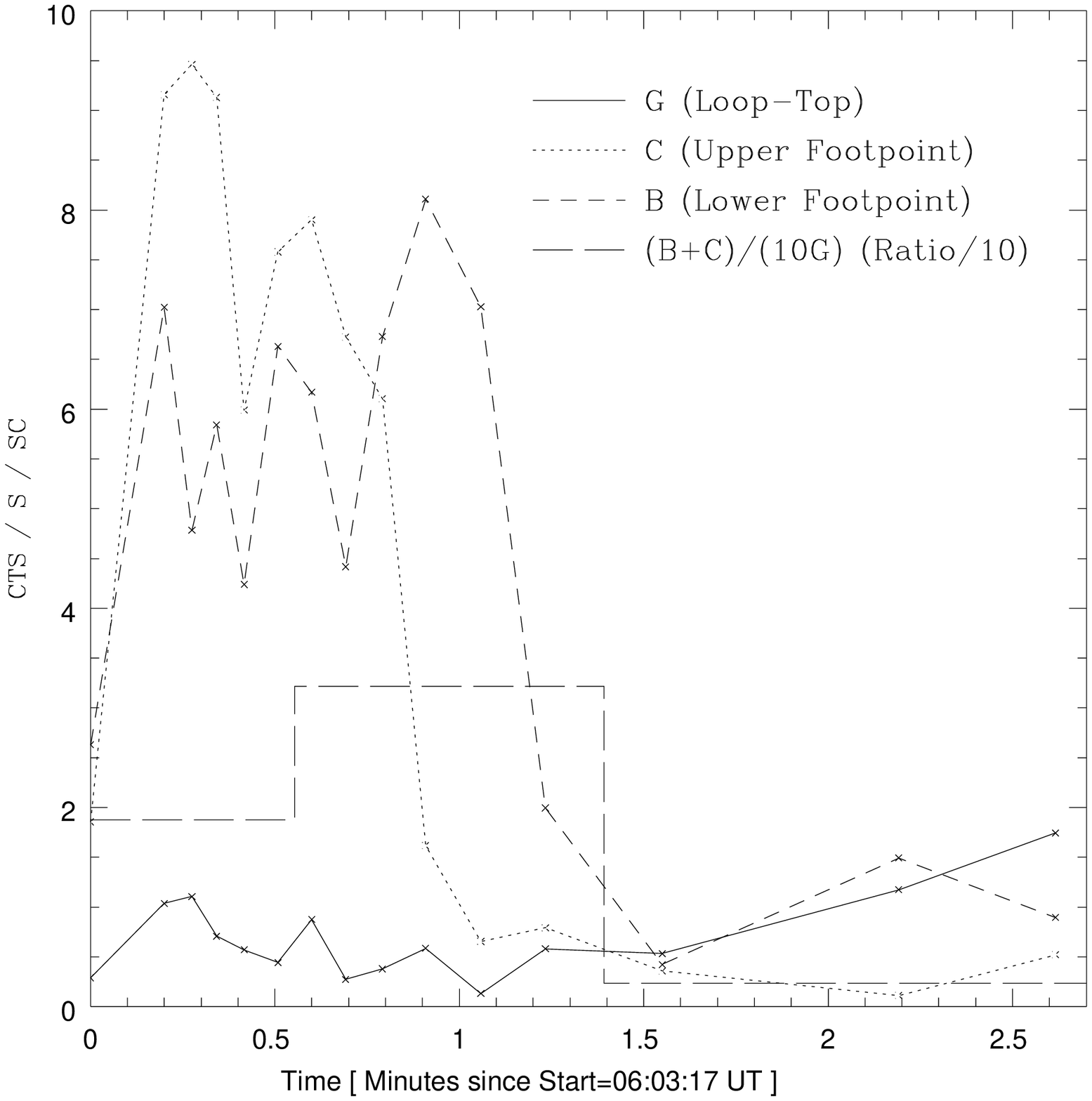,width=3in,height=2.6in}
\epsfig{file=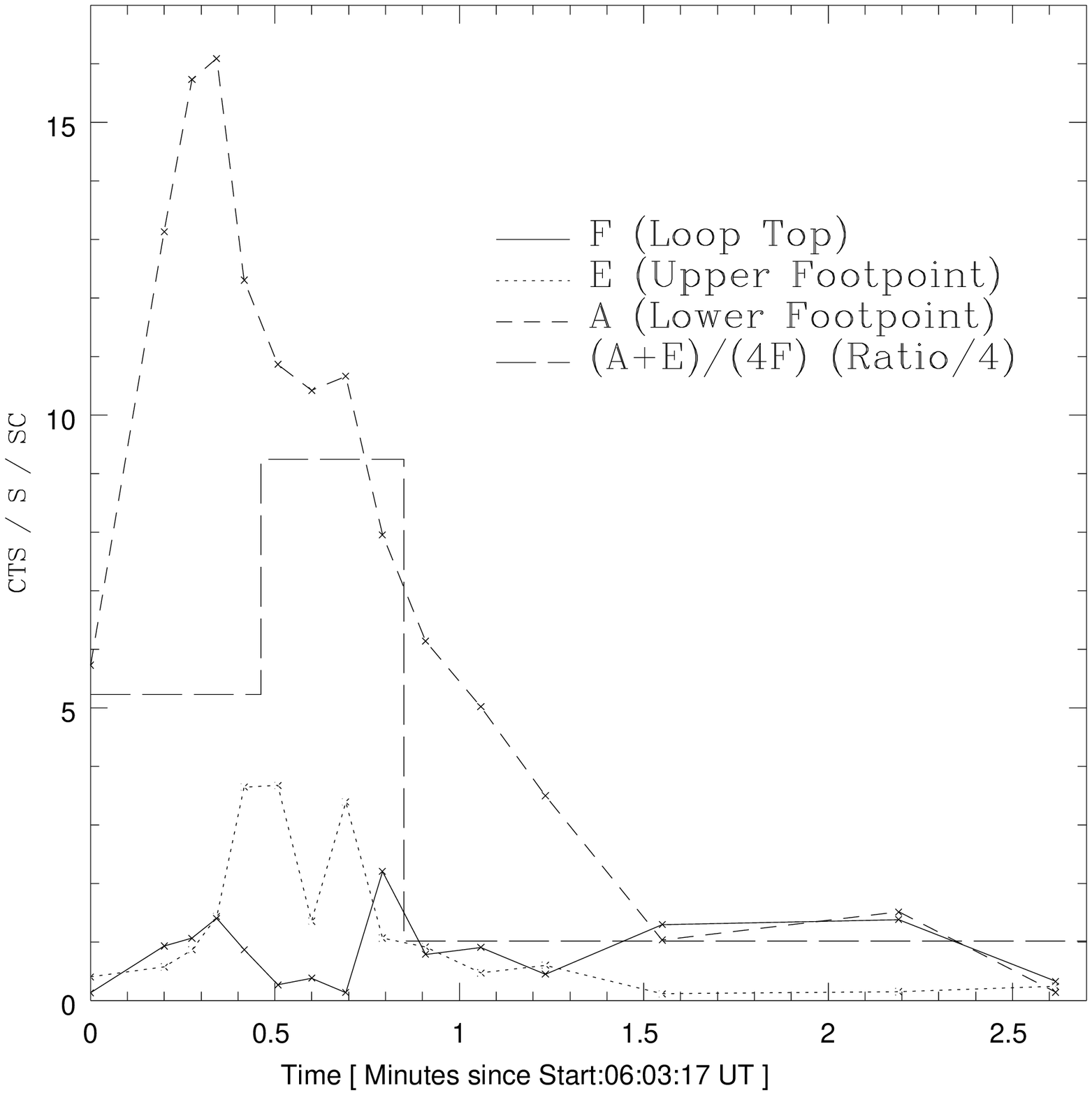,width=3in,height=2.6in}\vspace{-0.4cm}
\epsfig{file=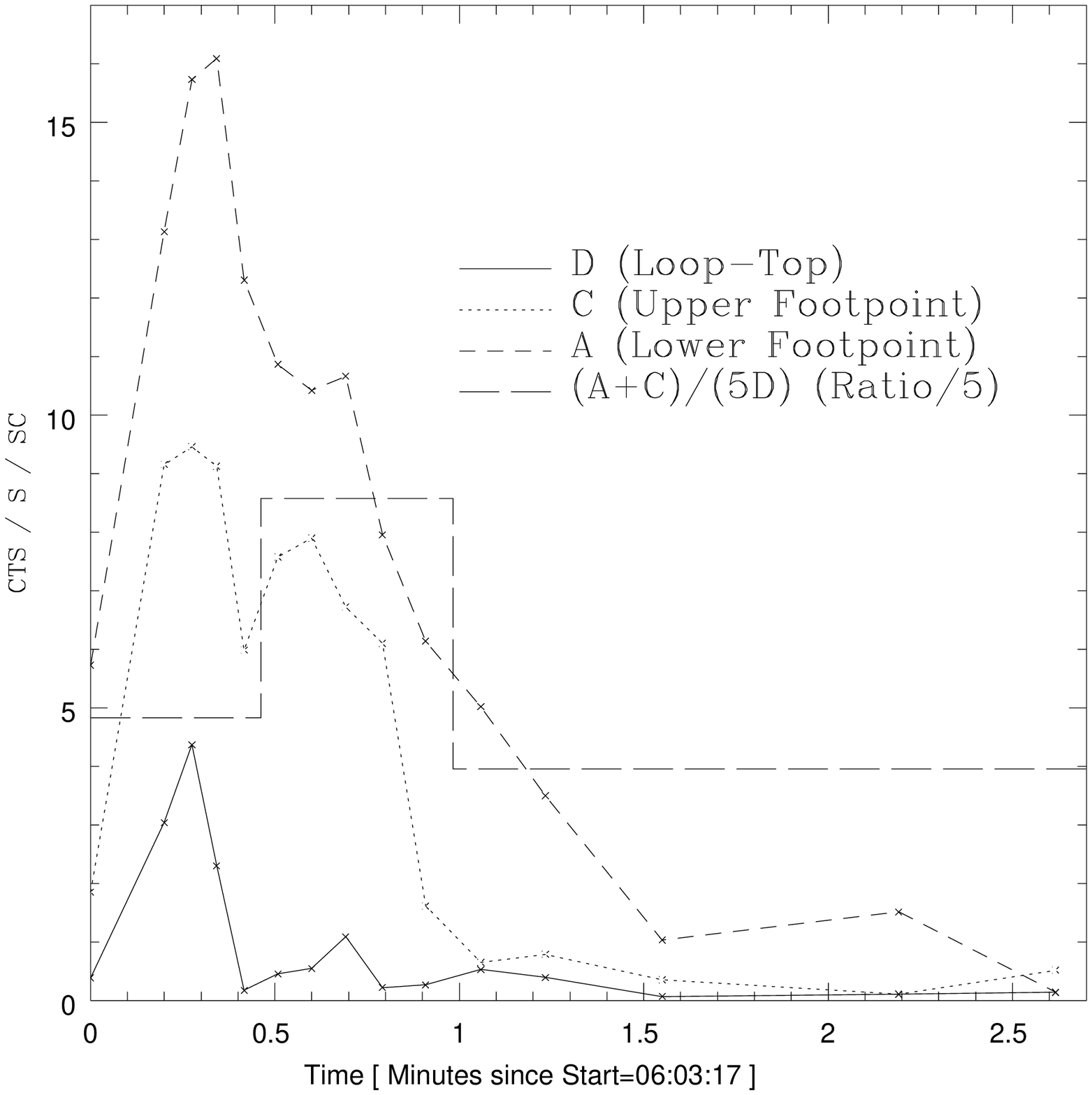,width=3in,height=2.6in}\hspace{1.1cm}
\psfig{file=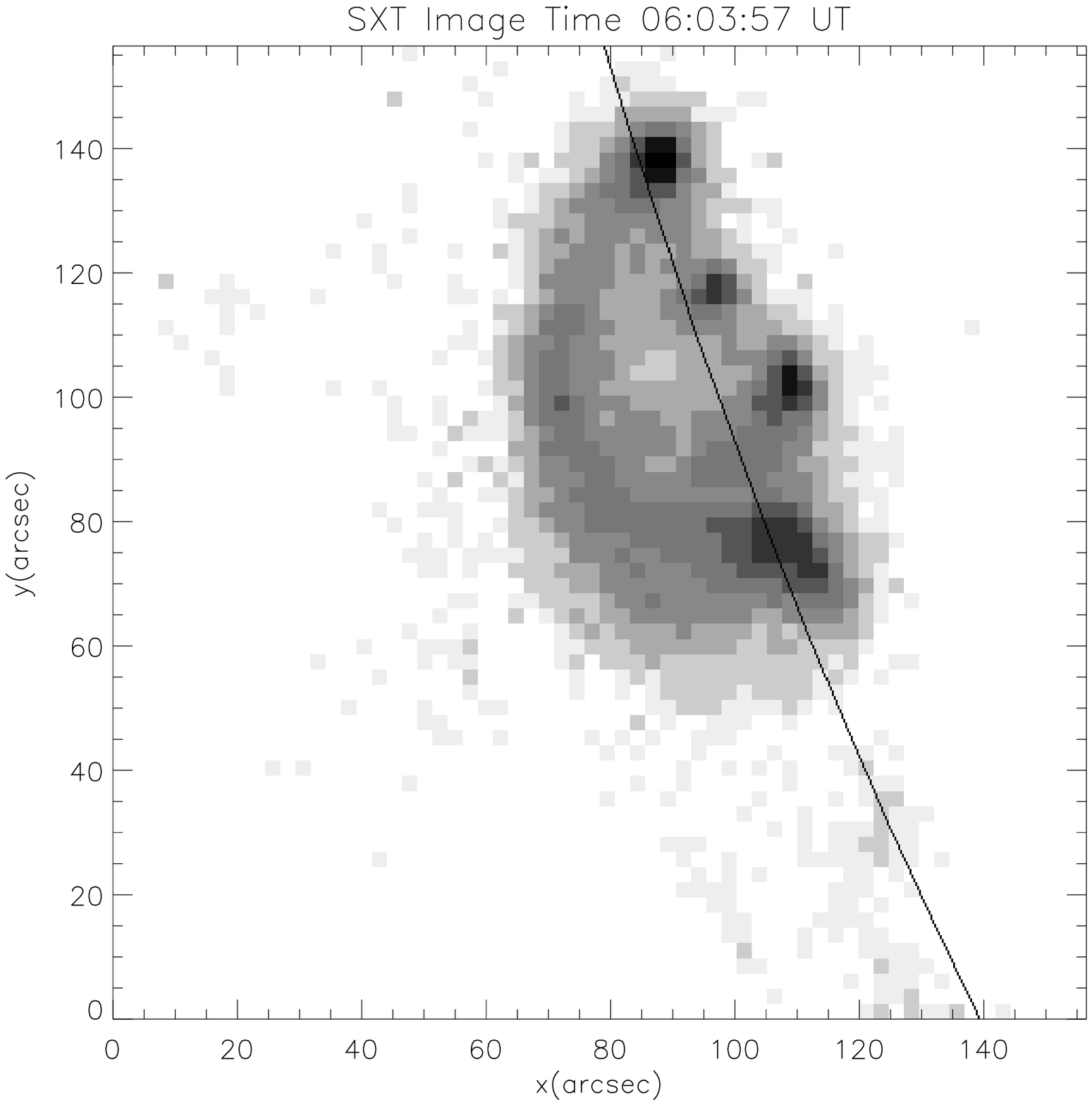,width=2.45in,height=3.88in,angle=90}
\psfig{file=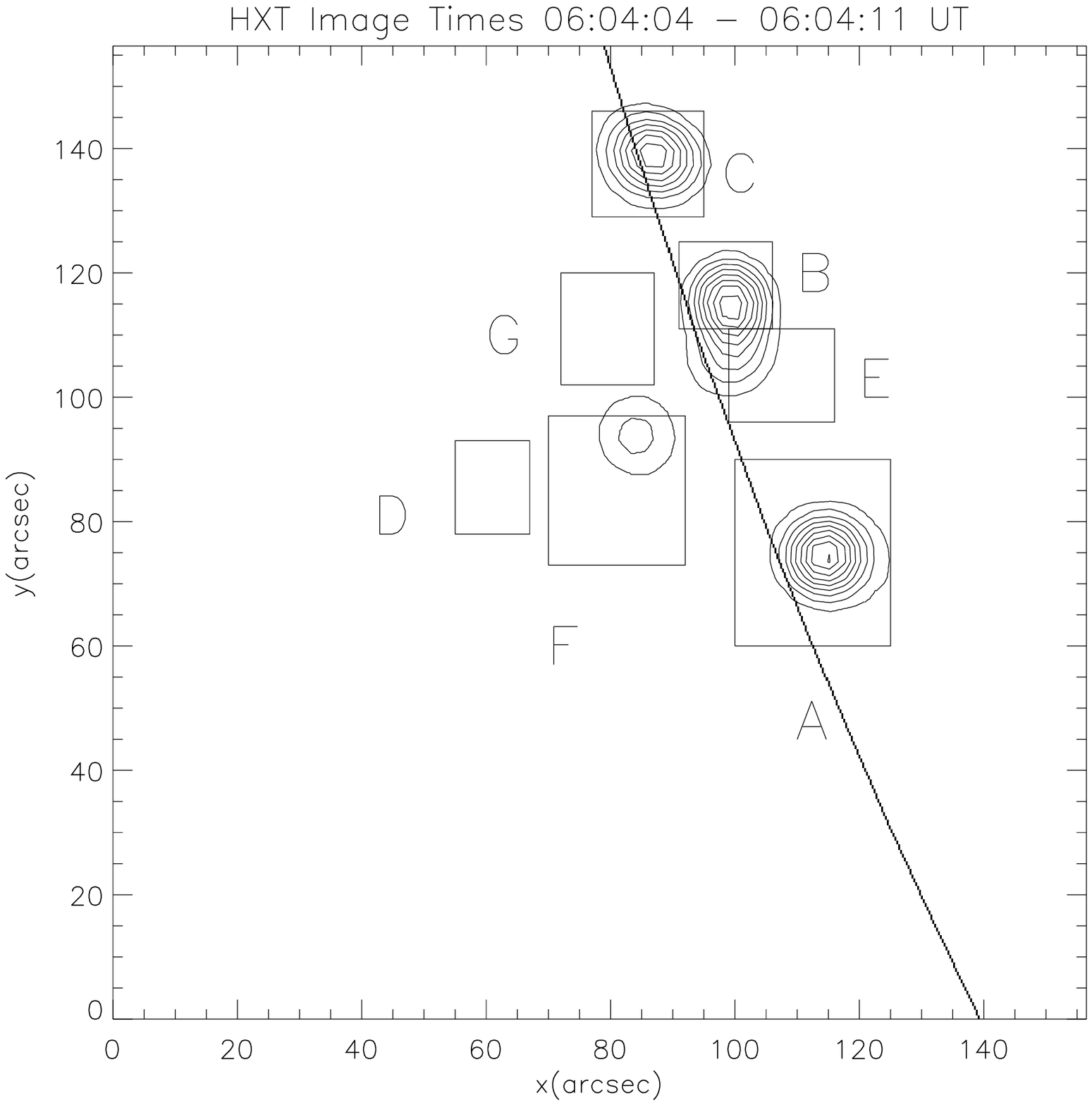,width=2.9in,angle=90}\hspace{-1.05cm}
\psfig{file=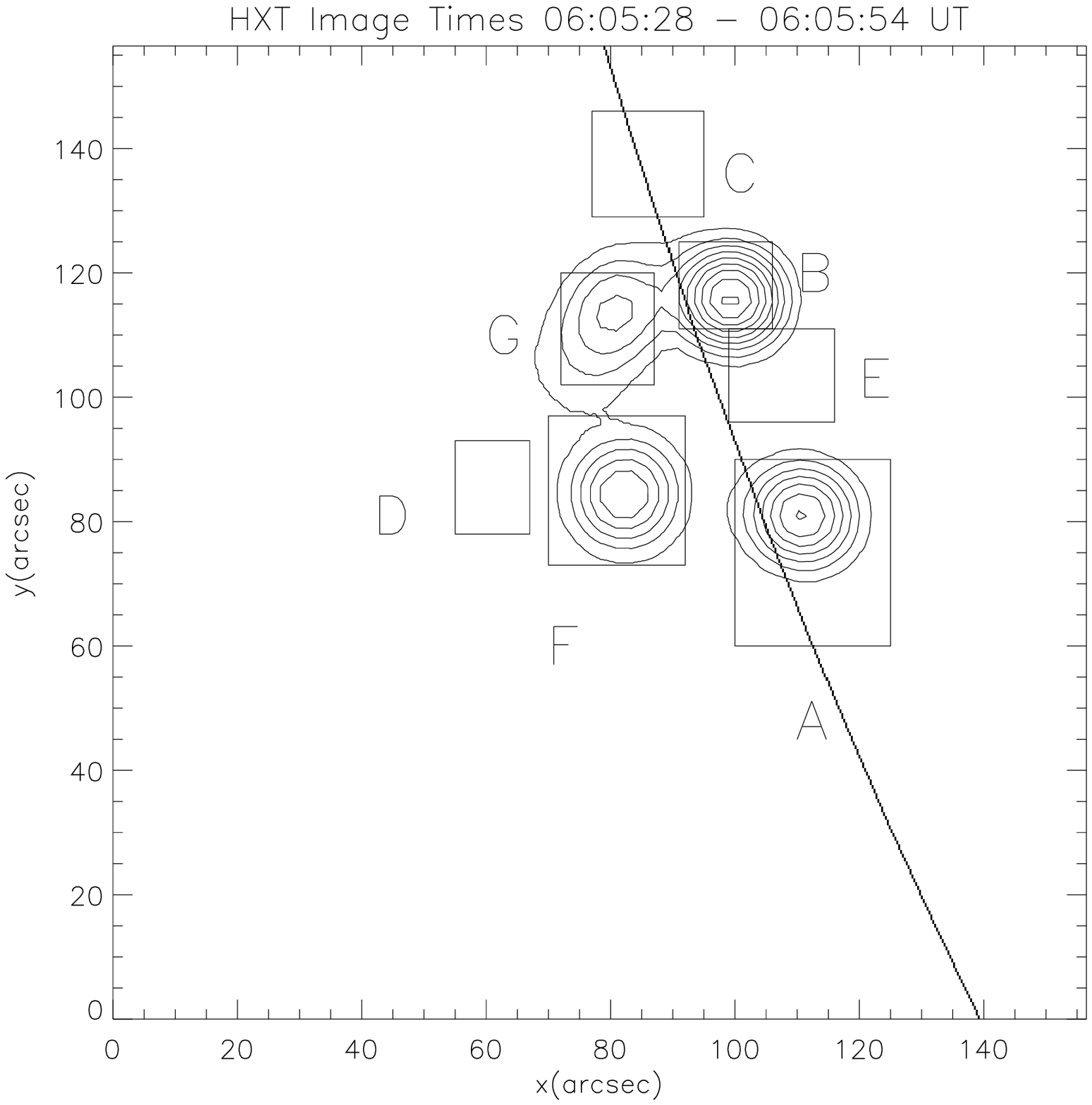,width=2.9in,angle=90}
\caption{Same as Figure \ref{b980818} for the November 30, 1993 flare.
The light curves in the middle left panel represents what may be the outer 
flaring 
loop (ADC), while those in
the upper two panels represent other sources seen in the flare, which may form 
two inner loops (CGB and EFA). The SXT image is shown in middle right separately 
and two HXT 
images at two different times are shown at the lower panels, with $B_{max}=2.5$  
and $\Delta B=0.23$ counts/pixel for the left panel, and $B_{max}=1.7$ and 
$\Delta B=0.16$ counts/pixel for the right panel.
These images were reconstructed using the ``pixon'' method.}
\label{931130}
\end{figure}

This flare was analyzed by Aschwanden et al. (1999) in their study of 
quadrupolar magnetic reconnection events. Aschwanden et al. analysis shows 
presence of only a double loop with the 
lower FP common to both the larger outer loop and the smaller inner 
loop.
Our 
analysis of this flare using the ``pixon'' method
resulted in images that were considerably cleaner than those from the  MEM 
technique. In Figure \ref{931130} we present a map of the X-rays images as well 
as 
light curves for the selected sources obtained from the ``pixon'' method. Based 
on these one may deduce that there are three separate loops BGC, AFE and ADC 
with common FPs A and C having composite time profiles. 

However, the basic 
structure of the flare remains complicated.
In the early stages of the flare there are actually four FP HXT sources 
that correspond to four bright spots in the SXT images (sources A, B, C \& E in 
the figure).
More confusing is the behavior of what appear to be three separate sources 
appearing high in the corona (sources D, F \& G).
The fainter source (D) appears early in the flare at the very top of the loop, 
while the two stronger sources (F \& G) appear later and at lower altitudes.
As the flare evolves, all of these FP and LT sources shift their 
position with respect to each other, making it very difficult to perform the 
usual analysis.
A comparison of the light curves for sources A, C \& D (the outer loop), shows 
that they peak at similar times early in the impulsive phase.
However, the peaks of the light curves for other sources are not correlated as 
clearly, and have extremely complicated structures. 
For example, a late peak of source B does not appear in any other source. Note 
also that the LT sources in this flare are much weaker than the FP sources.

\subsection{Miscellany}

\begin{figure}[htb]
\epsfig{file=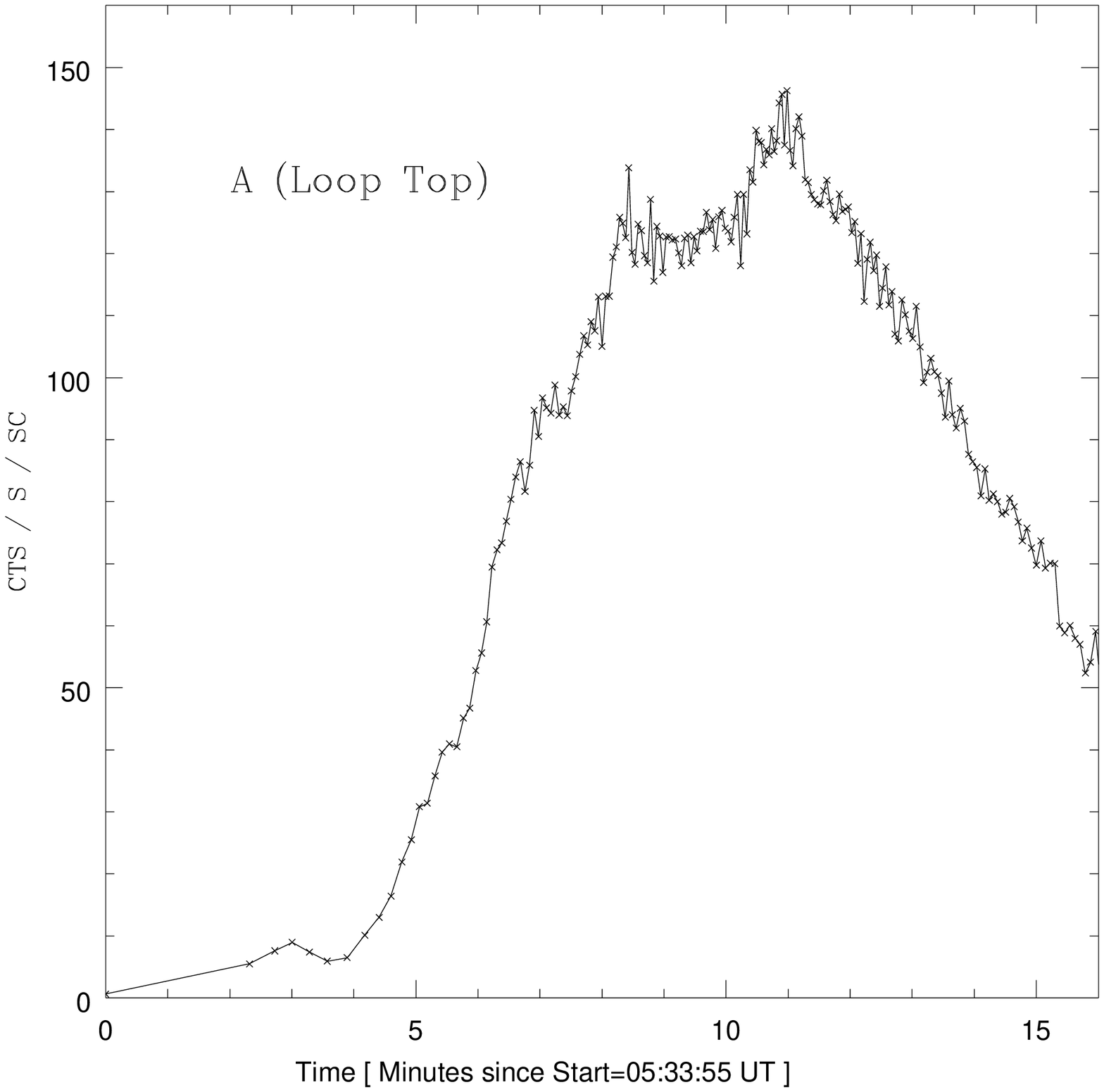,height=3in}
\psfig{file=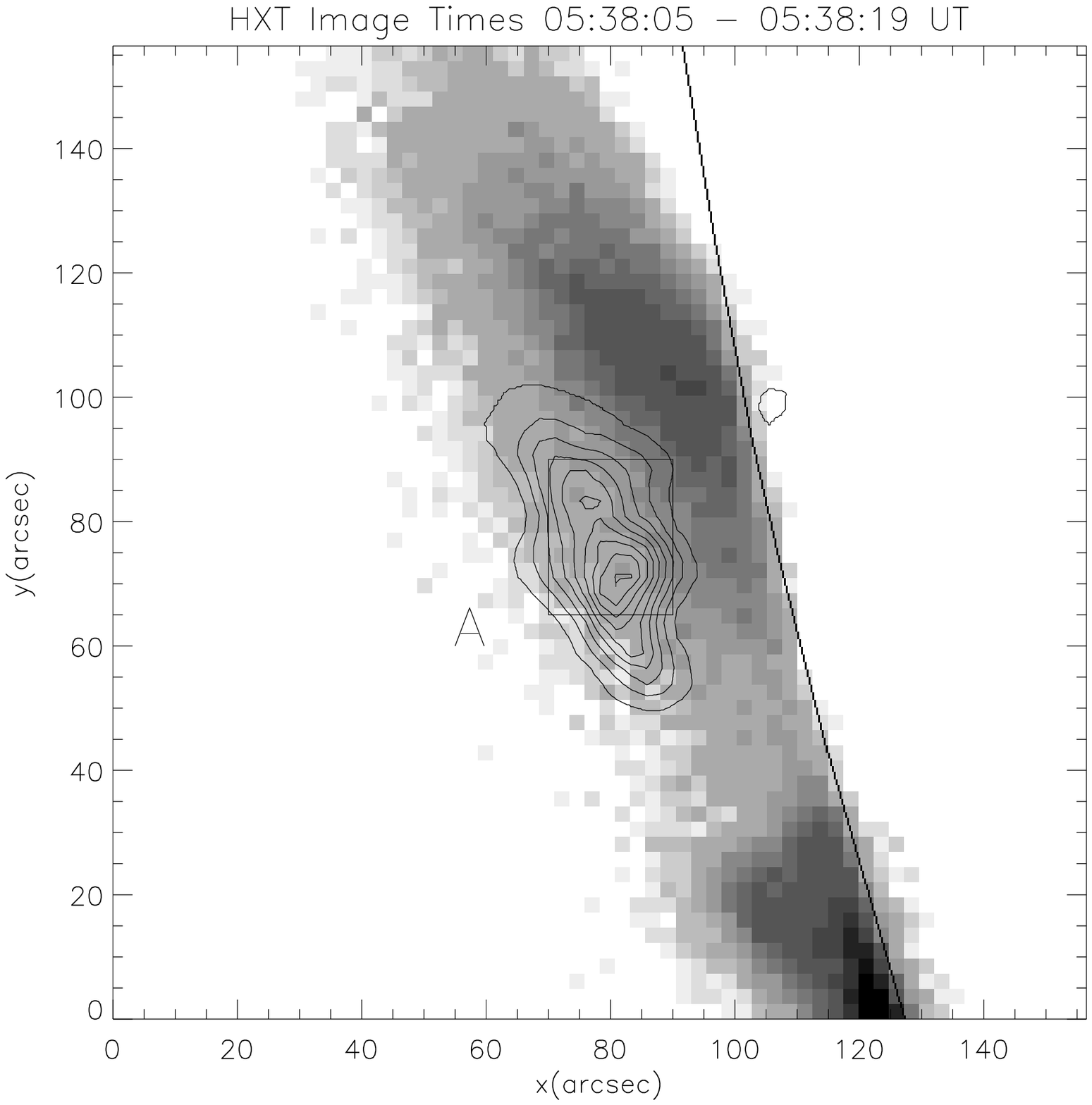,width=2.9in,height=4in,angle=90}
\caption{Same as Figure \ref{911218} for the April 23, 1998 flare, with 
$B_{max}=4.4$ and  $\Delta B=0.40$ counts/pixel. The FP 
sources of this flare are most likely occulted and source  A represents the LT 
source.}
\label{980423}
%
\epsfig{file=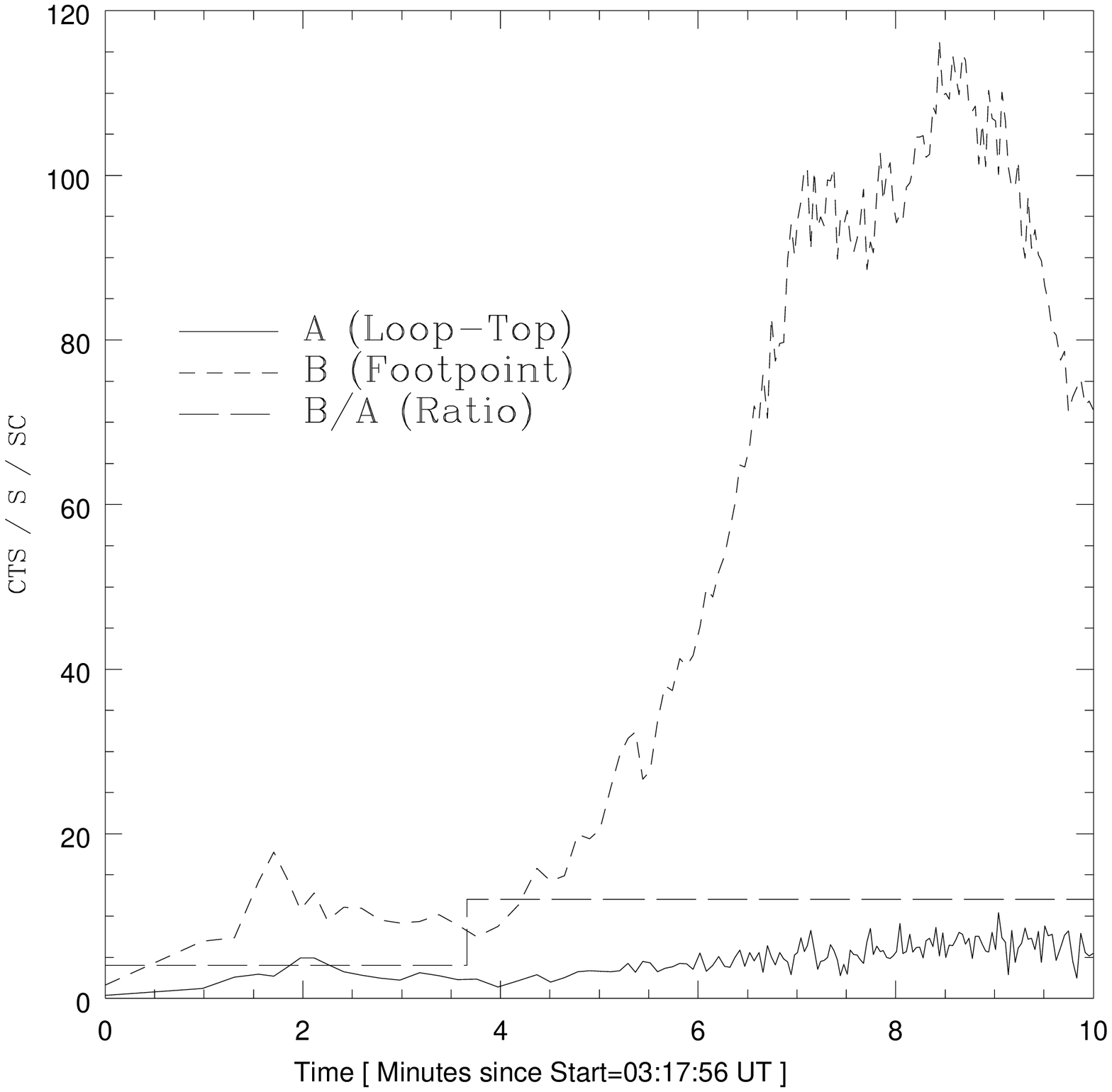,height=3in}
\psfig{file=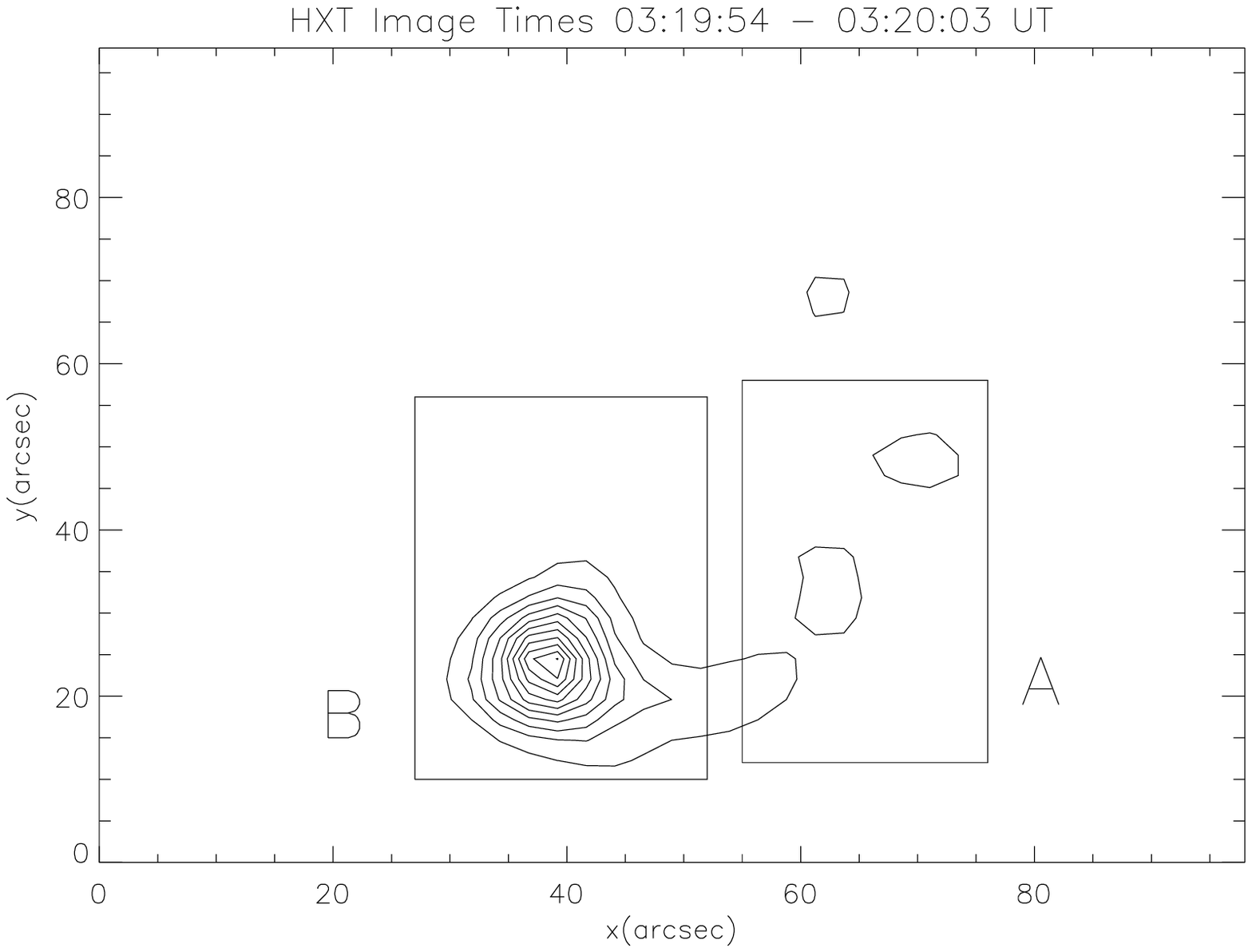,width=3.1in,height=2.9in}
\caption{Same as Figure \ref{911218} for the May 9, 1998 flare, with 
$B_{max}=4.1$ and $\Delta B=0.31$ counts/pixel.}
\label{980509}
\end{figure}

{\it April 23, 1998} -- This flare appears in the soft X-ray images as bright 
diffuse emission above the solar limb, accompanied by thin, quickly evolving 
loop structures that are probably not the main flaring loop.
The HXT emission initially appears in two sources, but quickly become one 
undifferentiated source associated with the bright diffuse soft X-ray emission.
The lack of an observable flaring loop indicates the possibility that this flare 
occurred behind the limb and the emission we are seeing is very high up in the 
corona, rather than a single FP source (see Figure \ref{980423}).

{\it May 9, 1998} -- This flare appears as a single FP and a LT 
hard X-ray 
source only during the initial two minutes of the flare (see Figure 
\ref{980509}),
when we see one strong FP accompanied by much 
fainter emission above it that may be a LT source.
As the flare progresses, the FP source grows in intensity so much that 
the limited dynamic range of the HXT makes the 
detection of the much fainter LT signal impossible.
The discrepancy between the positions of the SXT and HXT images is large, so 
that it is difficult to determine whether the observed HXT sources 
are associated with any SXT features.

\section{STATISTICAL RESULTS} 

In this section we consider the statistical properties of the  flares listed in 
Table \ref{events}
and summarize the differences between the original Masuda flares and the 
new events classified by us.
In total there are 18 events for which we have sufficient data. Of these 15 show 
at 
least one impulsive source which can be classified as a LT source.
Most of these can be classified as nonthermal, and one (the 920206 flare) is 
classified by Masuda 
as a ``super hot'' thermal LT source.
The remaining three show no detectable LT emission.
As we shall see below these non-detections are due to the shortcomings of the 
instrument indicating that LT emission is a general 
feature of solar flares.
We discuss the statistics of two physical characteristics, namely the relative 
fluxes and spectral indices of the LT and FP sources.

\subsection{Relative Fluxes}

The most directly available data are the statistics of the relative values and 
distributions 
of the fluxes from the LT and FP sources.
Theoretically, the relative values of these emissions are important for 
determination of the characteristics of the models.
As shown in {\bf PD}, the ratio of LT to FP emission can be related 
to the ratio of acceleration and diffusion rates of the electrons as well as the 
plasma density and magnetic field strength.
Here, we examine the relative values of the LT and FP emissions for 
the entire sample.

\begin{figure}
\epsfig{file=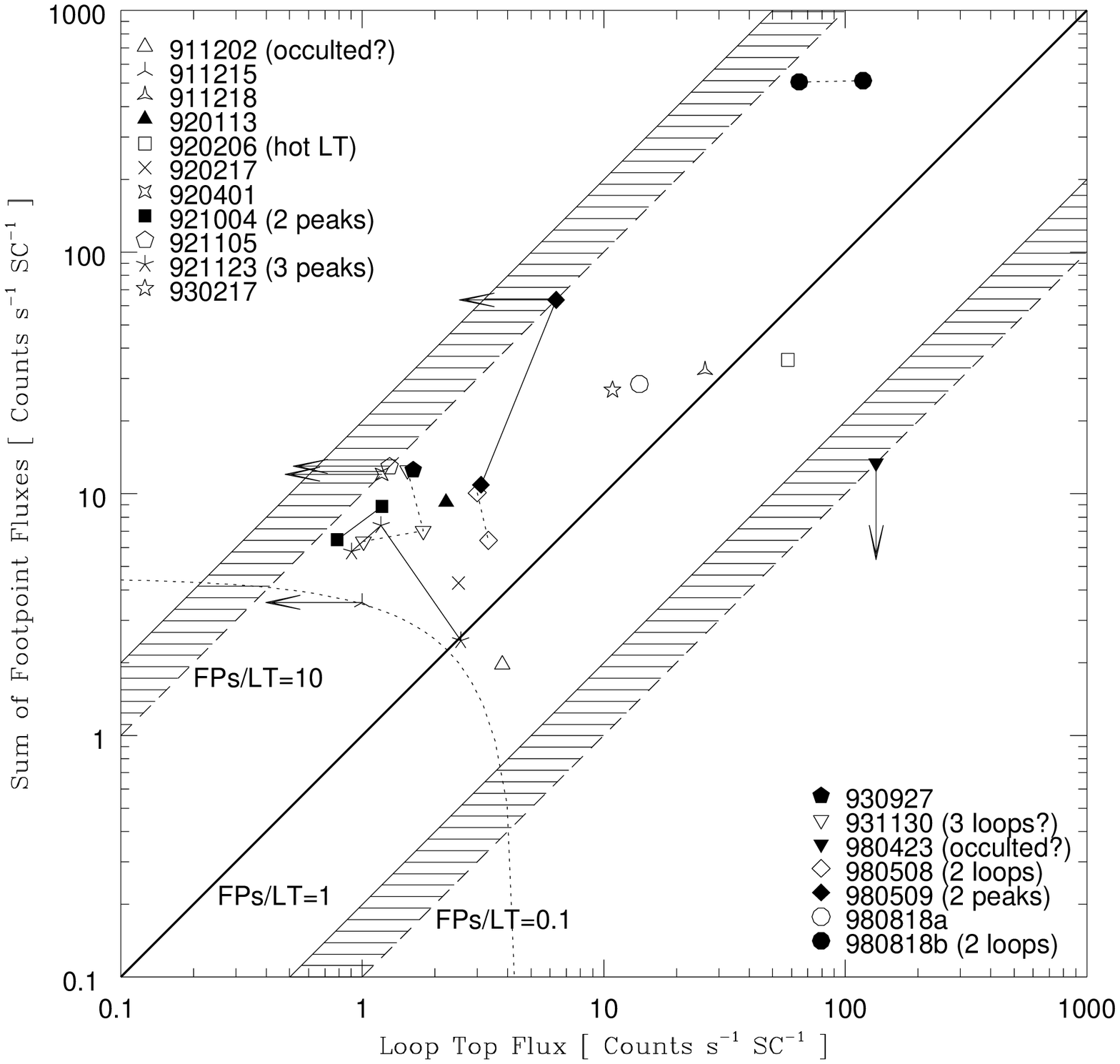,height=6in,width=6in}
\caption{Counts from two FPs vs. LT counts, in the M1 band  for all 
18 flares in our sample.
The diagonals represent lines of constant ratio (${\cal R}=FPs/LT$) and 
represent 
detection thresholds arising from the finite dynamic range of the instrument: 
Solid lines at 20 and 0.2 for two equal FPs; dotted diagonals 
for one dominant FP source.
Flares with undetected LT source are denoted by an arrow placed on 
the upper bound of detection of ${\cal R}=10$.
The dashed line shows the event selection threshold of 10 cts s$^{-1}$ SC$^{-1}$ 
at M2 band 
which on the average (see Fig. [\ref{stats}]) means $FPs+LT\simeq 5$ cts 
s$^{-1}$ SC$^{-1}$ 
for the M1 band.
For uniformity, all the data points on this plot were obtained using the 
standard MEM reconstruction procedure.}
\label{fluxes}
\end{figure}

In Figure \ref{fluxes}, we plot the sum of the count rates (in units of 
cts s$^{-1}$ SC$^{-1}$) of the two footpoint sources ($FPs$)  vs. the count 
rates of 
the corresponding looptop source ($LT$).
The values shown in this figure  refer to count rates (or fluxes) spatially 
integrated  
over each source, and integrated  over the entire impulsive 
duration of each flare, or over individual pulses in flares where distinct 
pulses and loops are discernible.
For three events, where multiple loop structures  with corresponding LT 
and FP sources are observed (see \S 3), we plot LT and FPs fluxes for each loop 
as separate points  but connect them with lines.

The primary trend seen in this figure is an obvious correlation between FPs and
LT fluxes over more than two decades of flux.  The second striking feature is
that for most flares (or pulses in flares) the ratio of these two fluxes is
confined to $1<{\cal R}=FPs/LT<10$.  Only three events lie beneath the line of
equality (the thick solid line).  The presence of an upper envelope is most
likely an artifact of the limited dynamic range of the instrument and the image
reconstruction process (estimated to be about one decade), and is not an
intrinsic feature of the flare emission.  This limitation means that for two FPs
with equal fluxes the plotted ratios will lie in the range $20<{\cal R}<1/5$
shown by the two thin solid diagonal lines.  But if one FP is much stronger than
the other, then the above ratio should lie in the range $10<{\cal R}<1/10$,
shown by the dotted lines.  (Because of the limited dynamic range, the weaker FP
source cannot be more than ten times weaker than the stronger one.  Hence,
strictly speaking the latter limits should be 11 and 0.11.)  Since the LT source
that is 10 or 20 times weaker than the accompanying FPs would not be detected by
the HXT instrument, or revealed by the image reconstruction process, we show the
three events with no detectable LT emission with horizontal arrows starting at
the ${\cal R}=10$ line, which assumes equal counts for the two FPs.  Similarly,
events falling beneath the ${\cal R}=0.1$ (or 0.2) line will not have a
detectable FP emission.  There is only one flare with the ratio approaching this
value.  This is the April, 24 1998 flare, shown by a vertical arrow in Figure
\ref{fluxes}.  This and one of the other two events below ${\cal R}= 1$ line,
the December 2, 1991 flare, are believed to have occurred behind the limb, so
that the FP sources are fully or partially occulted, giving it an abnormally low
flux ratio.  (Masuda mentions that two other events, January 13, 1992 and April
1, 1992, may also have occurred behind the limb.)  The other flare below the
${\cal R}=1$ is the February 6, 1992 event, which according to Masuda shows a
(super-hot) thermal spectrum for the LT and FP sources.  Our spectral analysis
shows a very steep spectrum for these sources (see Table \ref{3} below)
supporting this assertion.  The Dec.  18 1991 event, missed by Masuda, also
shows a very steep spectrum for both LT and FP sources (see Table \ref{3} below)
and lies near the equality line ${\cal R}=1$.

We therefore conclude that the absence of events below the line of equality
(${\cal R}=1$) is not an instrumental effect and must be intrinsic to the flare
process.  It should be reemphasized that the limitations introduced by the
dynamic range will also bias the data in favor of events with FP flux ratios of
less than 10.  Those with greater ratio will appear as a double (FP and LT) or
just a single (FP) source depending on the strength of the LT source.  We also
note that for some, but not all, events with multiple pulses and or multiple
loops, the data points from different loops or pulses lie near each other (e.g.
October 4, 1992 and May 8, 1998).  Interesting exceptions are the flare on
November 23, 1992, which exhibits one pulse with a significantly lower value of
${\cal R}$ than the other two pulses, and the flare on May 9, 1998, where a LT
source is detected during the first pulse, but not during the later, much
stronger pulse (which we show in Fig.  \ref{fluxes} with an upper bound for the
LT source).

We must also consider the truncation 
introduced by our selection criteria.  Because of the threshold of 10 
cts s$^{-1}$ SC$^{-1}$ imposed by us (and Masuda), only flares with the flux (in 
M2 band) of 
$(FPs + LT) > 10$ cts s$^{-1}$ SC$^{-1}$ 
will be in the sample. We estimate below that this translates to a threshold of 
5 cts s$^{-1}$ SC$^{-1}$ for the M1 band. This truncation 
boundary is shown by the dashed curve in Figure \ref{fluxes}. As is evident, all 
flares 
except the 911215 flare fall above this boundary.

\begin{figure}
\epsfig{file=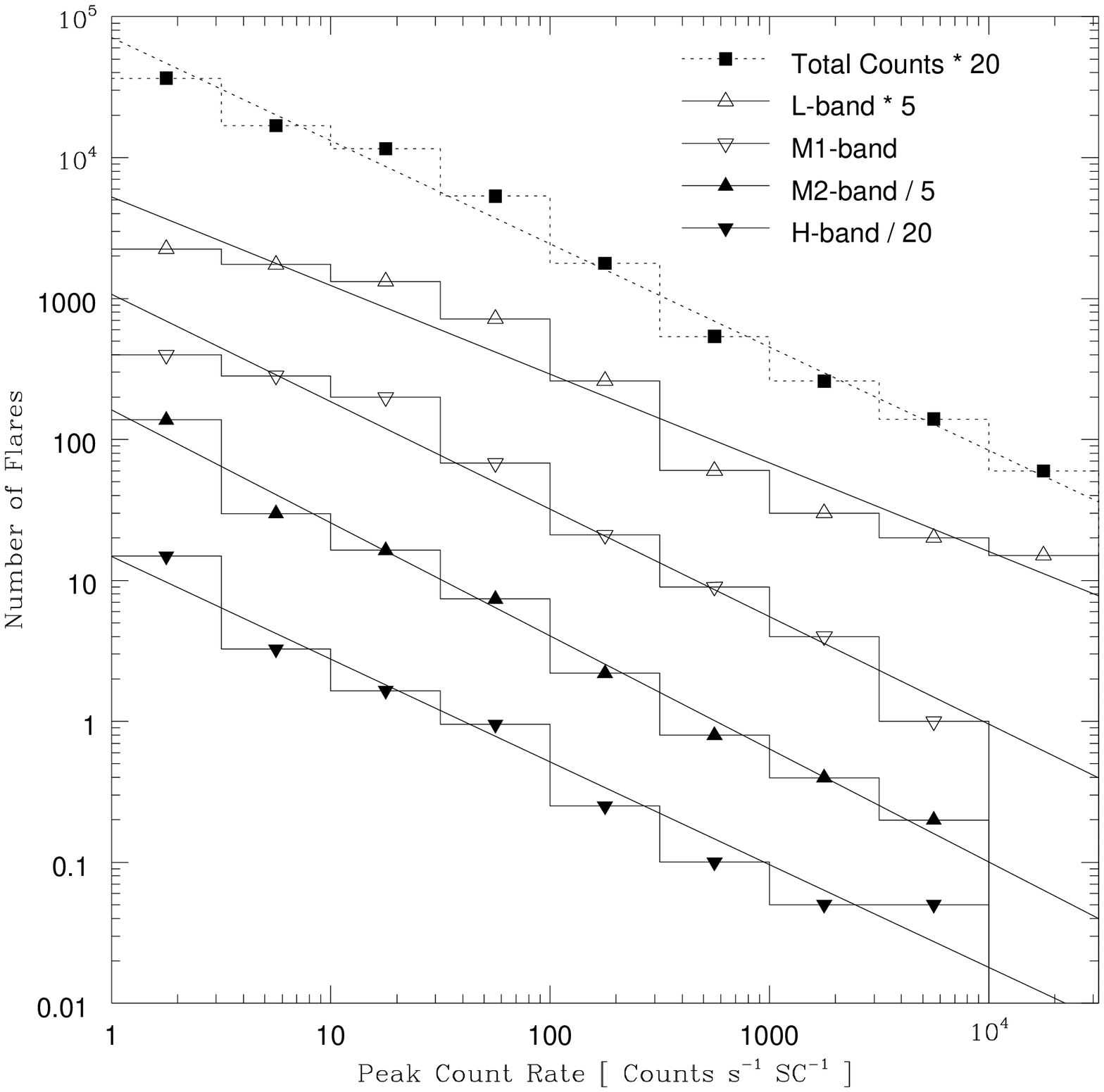,height=6in,width=6in}
\caption{The differential distribution of peak counts of all the flares  for 
the 
entire mission (October 1991-August 1998), for the sum and each of the four 
channels (bands 
L, M1, M2 and H, from top to bottom)  of the HXT instrument. For clarity the 
histograms  are shifted vertically by the 
indicated factors.
The straight lines show the linear regression fit for each histogram. The values 
of the  slopes  $\gamma$ are -0.73, -0.63 ,-0.76, -0.80 and -0.73 from top to 
bottom, respectively.}
\label{stats}
\end{figure}

In addition to the bounds derived from our selection criterion and the finite
dynamic range, there is also an absolute bound below which the instrument
sensitivity drops rapidly.  Since the HXT detection of flux is not based in a
well defined trigger process, we attempt to determine this bound by studying the
distribution of flare peak count rates using the entire HXT database of 1307
flares over the same time period.  In Figure \ref{stats}, we plot the
differential distribution of flare peak count rates for each of the HXT four
channels and for the summed over four channel peak count rate.  From each count
rate we subtract a representative background count rate for the channel.  Sato
et al.  (1998) give approximate background count rates of $L\simeq 2$,
$M1=M2\simeq 1$, $H\simeq 9$ cts s$^{-1}$ SC$^{-1}$.  As is evident, the 
distributions roughly
follow a power law, especially at high and midrange values, in agreement with
many previous flare surveys (see Lee \& Petrosian 1995 and reference cited
there).  We also plot the power law distributions from linear regression fits to
the distributions from M1, M2 and total peak count rates and give the power-law
indices ({\it i.e.}  logarithmic slopes).  The slope values of $\sim -0.6-0.8$ 
obtained here are consistent with previous results (Lee \& Petrosian 1995)
and shows that YOHKOH flares do not suffer from additional biases and that our
results regarding the LT emission is applicable generally to all flares.  The 
flattening away from the power law at low
count rates is due to a decrease in the sensitivity of the
detectors. This is especially prominent in the L channel, because of its
higher background rates.  For the M1 and M2 bands there seems to be very little 
flattening
down to the quoted background value of 1 cts s$^{-1}$ SC$^{-1}$.  Note that the 
M2 and M1 band
distributions are shifted relative to each other horizontally (and vertically
too) by an approximately constant value of about 2.  This means that the average
threshold value of the M1 cts s$^{-1}$ SC$^{-1}$ for our sample is about 5.  
This value was
used in determining the truncation boundary in Figure \ref{fluxes} (the dashed
line).

\begin{figure}
\epsfig{file=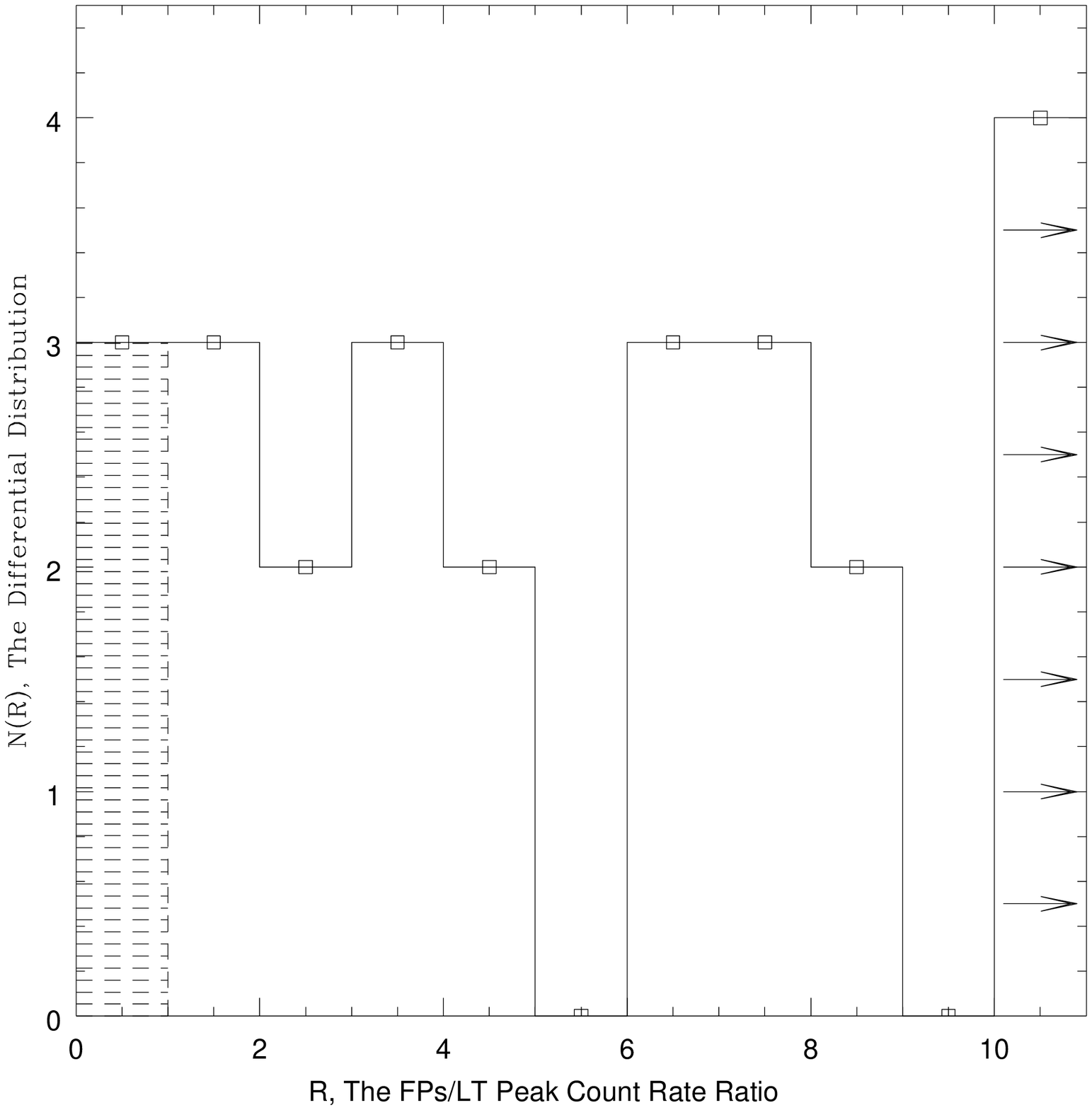,height=6in,width=6in}
\caption{The differential distribution of the ratio ${\cal R}=FPs/LT$ of the 
footpoints to looptop peak counts of all the flares  for 
the 
entire mission (October 1991-August 1998). The arrows indicate ratios greater 
than the dynamic range of about 10. Some of the flares in the shaded area may be 
occulted or be dominated by thermal emission from a superhot component.}
\label{ratio}
\end{figure}

In summary, from the above analysis of the relative strengths of the LT 
and FP 
fluxes we can  conclude that the hypothesis that all flares have LT 
emission is consistent with the extant, though limited, data.
 Figure 
\ref{ratio} shows the distribution of ${\cal R}=FPs/LT$ for the flares presented 
in Figure \ref{fluxes}. This distribution if fairly flat between   0.5 to 8 and 
there are
four flares with undetected LT emission ({\it i.e.}  ${\cal R} > 10$) as shown 
by 
the arrows. Obviously the statistical significance of these data point is 
marginal 
and a more quantitative account of the relative distributions of the fluxes of 
the 
LT and FP sources requires a larger sample and dynamic range. 
Nevertheless, this kind of data is very important and can constrain the model 
parameters. As  
described in {\bf PD}  this ratio can be related to the acceleration and the 
background plasma parameters. For example, it was shown that under certain 
conditions one has ${\cal R}= \tau_{\rm Coul}/T_{\rm esc}$, where $\tau_{\rm 
Coul}$ and $T_{\rm esc}$ are the Coulomb collision time scale in the loop and 
the escape time on the electrons from the acceleration site at the LT. However, 
for more general situation the relation of this ratio to the physical parameters 
si more complicated. As shown by numerical calculation in {\bf PD} values of a 
few for this ratio indicate plasma densities of about 
few times $10^{10}$ cm$^{-3}$ and require a relatively short acceleration time 
compared to 
the escape time of the electrons from the LT source. Further constraints 
come from the consideration of the spectral shapes which we discuss next.

\subsection{Spectral Indices}

Another characteristic of the LT and FP emissions which can shed light on the 
acceleration mechanism  is the relative shape of their spectra.
The HXT instrument  observes only at four broad channels, so that an accurate, 
absolute spectral analysis of individual sources is difficult and 
results are  uncertain. However, the relative spectral characteristics of 
FP and LT sources are more reliable.
To this end we have carried out the following analysis.
We use the YOHKOH routine \sf hxtbox\_fsp, \rm which fits a specified model 
spectrum to the observed count rates and  finds the 
best fit model parameters and calculates the chi-square's. When possible, we 
attempted to measure the spectral  shape early in the impulsive 
phase (before or right at the first impulsive peak) so as to minimize 
any thermal contributions to the spectrum.

For flare events with detectable images in the H band, we use the following 
four 
models:

1. A simple power law fitted to all four channels (Powerlaw1).

2. A simple power law fitted to the three highest channels (Powerlaw2) so that 
we avoid possible thermal contribution in the L band.

3. A broken power law  with three parameters (low and high energy spectral 
indices, $\gamma_1$ and $\gamma_2$, and the break energy $E_{br}$) fitted to all 
four channels (Broken Powerlaw1).

4. Two simple power laws fitted separately to the L and M1 bands, and to the M1 
and M2 bands (Broken Powerlaw2). In this case the break energy is fixed as the 
midpoint of the M1 band at $\sim 28$ keV.

Table \ref{2} summarizes these results.


\newcommand{\1}{\hspace{1mm}}
\newcommand{\2}{\hspace{2mm}}
\newcommand{\3}{\hspace{3mm}}
\newcommand{\x}{\hspace{9mm}}

\begin{table}[ht] \centering \scriptsize
\begin{tabular}{|c|l|l|l|l|l|l|l|}\hline\hline
Event \# & Date & Time & Source & Powerlaw1 & Powerlaw2 & Broken Powerlaw1 & 
Broken 
Powerlaw2 \\ 
&&&&$\1 \gamma$\3 $\chi^{2}/\nu$  &  \1 $\gamma$\3 $\chi^{2}/\nu$ 
	&   $\gamma_{1}$\3 $\gamma_{2}$\3 $E_{br}$\3 $\chi^{2}$ 
	&   $\gamma_{L,M1}$\3 $\gamma_{M1,M2}$ \\
\hline
2&911215 & 02:44:09 & A (FP) & 6.1\3 21 & 4.5\3 5.8  & --  &\1 6.8\x 3.8\\
& & & B (FP) & 4.3\3 18 & 5.1\3 16   & 3.0\3 9.9\3 42\3 0.0&\1 2.9\x 3.5 \\
\hline
4 & 920113 & 17:28:04 & A (FP) & 3.5\3 2.2  & 3.8\3 0.32  & -- &\1 3.1\x 3.9 \\
&       &          & B (FP)  & 3.3\3 8.6    & 3.7\3 0.02 & --  &\1 2.5\x 3.7 \\
&       &          & C (LT) & 5.3\3 12.8     & 6.6\3 0.45  & -- &\1 4.2\x 6.8 \\
\hline
7&920401&10:13:05&A (FP)&3.8\3 1.3&3.9\3 0.0&2.2\3 3.9\3 18\3 0.0&\1 3.7\x 3.9 
\\
\hline
8 & 921004 & 22:19:31 & A (FP) & 4.7\3 21  & 5.9\3 0.01  & --  &\1 3.2\x 5.9 \\

&  &     & B (FP) & 4.5\3 14   & 4.9\3 19    & --	       &\1 4.0\x 3.6 \\		
&  &     & C (LT)& 5.8\3 0.61& 5.6\3 0.85& 5.8\3 7.6\3 56\3 1.1&\1 6.0\x 5.3 \\
\hline
10&921105&06:19:20&A (FP)&4.1\3 20&4.4\3 8.2&3.6\3 5.0\3 34\3 0.0&\1 3.6\x 4.1 
\\
\hline
19 & 980818 & 08:19:13 & A (LT) & 7.4\3 32  & 7.4\3 64   & --    &\1 7.4\x 7.7 
\\
   &        &          & B (FP) & 6.0\3 260 & 4.3\3 0.41 &12\3 4.3\3 18\3 .45&\1 
6.6\x 4.2 \\
   &        &          & C (FP) & 5.1\3 140 & 6.2\3 0.02 & --    &\1 3.9\x 6.2 
\\
\hline 
20&980818&22:14:36&A (FP)&3.1\3 590&3.5\3 200&2.3\3 3.9\3 32\3 0.0 &\1 2.2\x 3.1 
\\
&      &          & B (FP) & 3.6\3 2.4   & 3.6\3 1.3   & --  &\1 3.7\x 3.5 \\
&      &   & C (FP) & 3.5\3 17&3.4\3 1.9&5.6\3 3.4\3 18\3 1.9&\1 3.7\x 3.4 \\
&      &   & D (LT) & 5.0\3 66& 4.4\3 59 &5.3\3 3.2\3 37\3 0.0&\1 5.3\x 4.9 \\
&      &  & E (LT)& 3.8\3 32& 4.0\3 43& 3.2\3 3.8\3 9.7\3 63  &\1 3.4\x 4.4 \\ 
\hline\hline
\end{tabular}
\caption{Spectral parameters from fits to 
four models of events with detectable H band  images: Powerlaw1 uses all four 
channels (degrees of freedom $\nu=2$); Powerlaw2 uses the three highest 
channels ($\nu=1$); Broken Powerlaw1 uses all 4 channels ($E_{br}$ in units of 
keV); and 
Broken Powerlaw2 the three lowest 
bands with $E_{br}$  set to $\sim 28$ keV, the midpoint of the M1 channel 
($\chi^2=0, \nu=0$). Missing spectral parameters for events in the Broken 
Powerlaw1 column indicate
that the $\chi^{2}$ minimization for the spectra did not converge.}
\label{2} 
\end{table}

 
We  first note that in 14 out of 18 cases the value of the reduced  chi-square, 
$\chi^{2}/\nu$, is lowered when we remove the L band  from the fit. The 
4-channel fit chi-squares are lower, by statistically insignificant 
amounts, only for one FP and three LT 
sources. This combined with the results from the Broken Powerlaw2 fits indicates 
that deviations from a power law  is produced by the L band and that these 
could 
be in form of a flattening or steepening of the spectrum  with approximately 
equal 
probability. (The break energy $E_{br}$ distribution, however, does not seem too 
agree with this conclusion. Only 4 out of nine values fall in or near the L 
band. 
This could be due to statistics of small sample.)  The observed spectral change, 
of course, is not a reflection of the true behavior of the 
spectra because flares with detectable H band flux must necessarily be biased 
toward those with flat and/or flattening spectra. Below we will compare the 
statistics of these flares with those without a detectable H band flux. 

For flares with no useable images in the H band , we use two models;
a simple power law  fitted to the three  channels (Powerlaw) and the model 4 
above (Powerlaw2). Table 
\ref{3}  presents the results of this analysis.

\begin{table}[ht] \centering \scriptsize
\begin{tabular}{|c|l|l|l|l|l|}\hline\hline
Event \# & Date & Time & Source & Powerlaw & Broken Powerlaw2 \\ 
 &    & hh:mm:ss  &        & \1 $\gamma$ \3 $\chi^{2}/\nu$ 
	& $\gamma_{L,M1}$\3 $\gamma_{M1,M2}$ \\ \hline
1&911202 & 04:53:45 & A (LT) & 6.2\3  0.6    &\1 6.3\x  5.9  \\
  &     &          & B (FP) & 5.8\3  3.3    &\1 5.2\x  7.1  \\
  &     &          & C (FP) & 5.4\3  0.01   &\1 5.4\x  5.3  \\ \hline
3&911218 & 10:27:16 & A (LT) & 7.1\3  87     &\1 6.5\x  9.0  \\
  &     &          & B (FP) & 6.2\3  46     &\1 5.8\x  7.1  \\
  &     &          & C (FP) & 8.3\3  31	   &\1 7.9\x  11   \\ \hline
5&920206 & 03:22:16 & A (LT) & 8.1\3  320    &\1 6.8\x  12   \\
  &     &          & B (FP) & 7.7\3  43     &\1 6.9\x  10   \\
  &     &          & C (FP) & 8.3\3  56     &\1 7.6\x  11   \\ \hline
6&920217 & 15:40:44 & A (FP) & 3.3\3  14     &\1 2.4\x  4.1  \\
  &     &          & B (LT) & 6.2\3  31     &\1 6.9\x  3.1  \\
  &     &          & C (FP) & 3.5\3  41     &\1 3.6\x  3.5  \\ \hline
11&921123 & 20:24:39 & A (FP) & 3.7\3  1.7    &\1 3.5\x  3.9  \\
  &     &          & B (LT) & 10\3\1 2.0    &\1 10\x   8.0  \\
  &     &          & C (FP) & 6.4\3  3.2    &\1 6.7\x  5.5  \\ \hline 
12&930217 & 10:36:13 & A (FP) & 4.8\3  37     &\1 4.3\x  5.6  \\
  &     &          & B (FP) & 6.5\3  42.    &\1 6.0\x  7.9  \\
  &     &          & C (LT) & 6.2\3  12     &\1 5.8\x  7.0  \\ \hline
13&930927 & 12:08:10 & A (LT) & 6.2\3  8.3    &\1 5.6\x  9.7  \\ \hline
  &     &          & B (FP) & 4.9\3  20     &\1 4.1\x  7.2  \\
  &     &          & C (FP) & 5.7\3  23     &\1 5.7\x  7.2  \\
14&931130 & 06:03:33 & A (FP) & 2.7\3  37     &\1 1.6\x  3.5  \\
  &     &          & B (FP) & 3.8\3  3.6    &\1 3.2\x  4.4  \\
  &     &          & C (FP) & 3.5\3  74     &\1 1.3\x  5.7  \\
  &     &          & D (LT) & 3.6\3  18     &\1 2.1\x  5.2  \\ \hline
15&980423 & 05:38:46 & A (LT) & 7.5\3  5.2	   &\1 7.4\x  7.8  \\ \hline
16&980508 & 01:58:29 & A (FP) & 4.7\3  35     &\1 4.7\x  4.5  \\
  &     &          & C (LT) & 5.8\3  22     &\1 5.0\x  7.5  \\
  &     & 01:57:28 & B (FP) & 3.9\3  4.0    &\1 3.7\x  4.2  \\
  &     &          & D (FP) & 4.2\3  19     &\1 3.3\x  5.5  \\
  &     &          & E (LT) & 5.5\3  22     &\1 4.8\x  7.0  \\ \hline
17&980509 & 03:19:36 & A (LT) & 6.6\3  5.2	   &\1 6.1\x  7.9  \\ 
   &    &          & B (FP) & 6.3\3  20	   &\1 5.9\x  7.3  \\
\hline\hline       
\end{tabular}
\caption{Spectral indices of events wit undetected H band  images
from a simple powerlaw fit to
three available bands ($\nu=1$) and from a broken power law fit,  where we fix 
the break 
energy to correspond to the  midpoint of M1 band  at $\sim 28$ keV ($\chi^2=0$, 
$\nu=0$).}
\label{3}
\end{table}

We first would like to note that the results from our analysis of the same  
events do not agree 
with Masuda's numbers. This could be either because the HXT grid 
response functions have been improved since Masuda did his analysis (see Sato, 
Kosugi and Makishima 1999), or
due to the differences in  the spatial 
boxes or time intervals used to calculate the indices. Masuda does not specify 
these data. However, the  general characteristics of the spectra (ranges, 
steepenings, etc.) of pre and post September 1993 bursts given in the above 
table  agree 
with the the characteristics  of ten pre September 1993 
ten spectra cited by Masuda. 

Second, we note that in contrast to the sample of Table 2, only two sources (LT 
sources in 920217 and 921123) in Table 3  
show a significant flattening. The rest maintain the power law form, or in the 
majority of cases 
steepen by a significant amount (an spectral index change of greater than 1); 
the spectral index changing by more than 3 in 
some sources. This difference between the flares from Tabels 2 and 3 is due to 
the selection bias mentioned above and 
indicates that caution should be exercised in the interpretation of results 
from 
limited samples. The following figures describe the prominent features of the 
results from the whole sample.

\begin{figure}
\epsfig{file=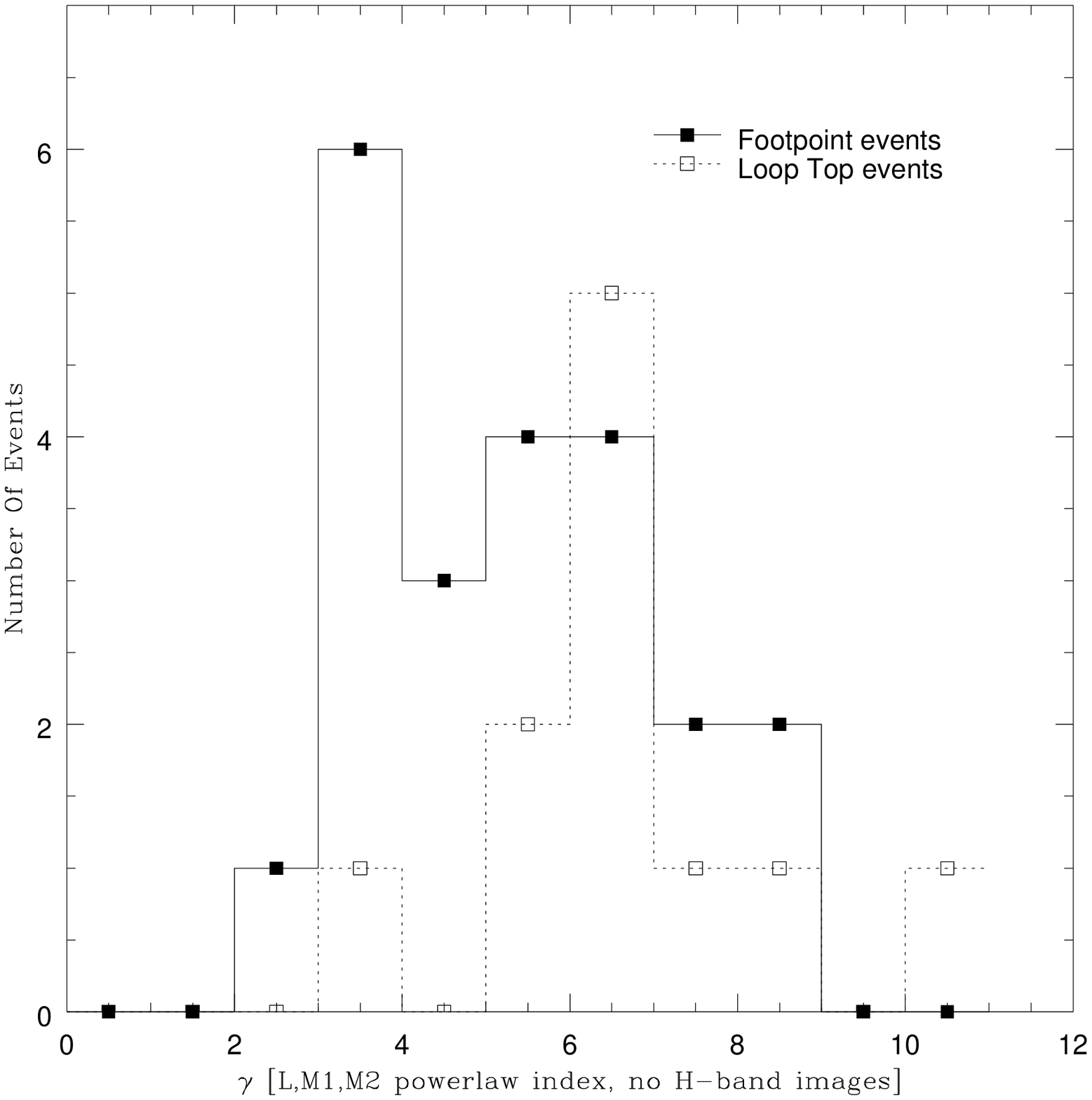,height=2.9in}
\epsfig{file=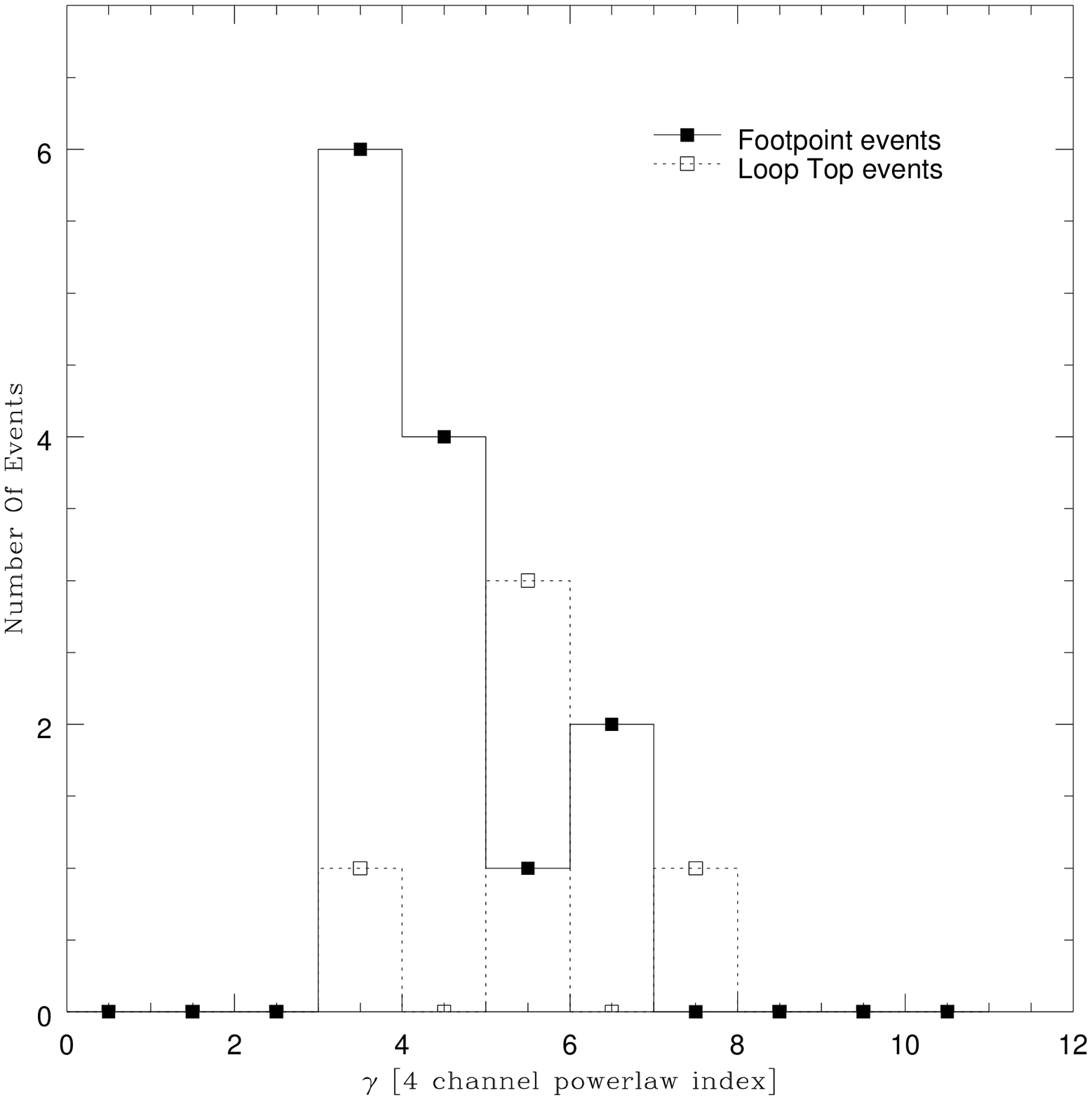,height=2.9in}
\caption{Distribution of power law  spectral 
indices: {\bf Left panel} from fits to the L, M1 and M2 bands for all events 
with no 
detectable H band  images; {\bf right panel} from fits to all 
four bands for events with detectable H band images.
The solid histograms and filled points represent the FP sources and the 
dotted histogram and open points represent the loop 
top sources.}
\label{3chindex}
\end{figure}

\begin{figure}
\epsfig{file=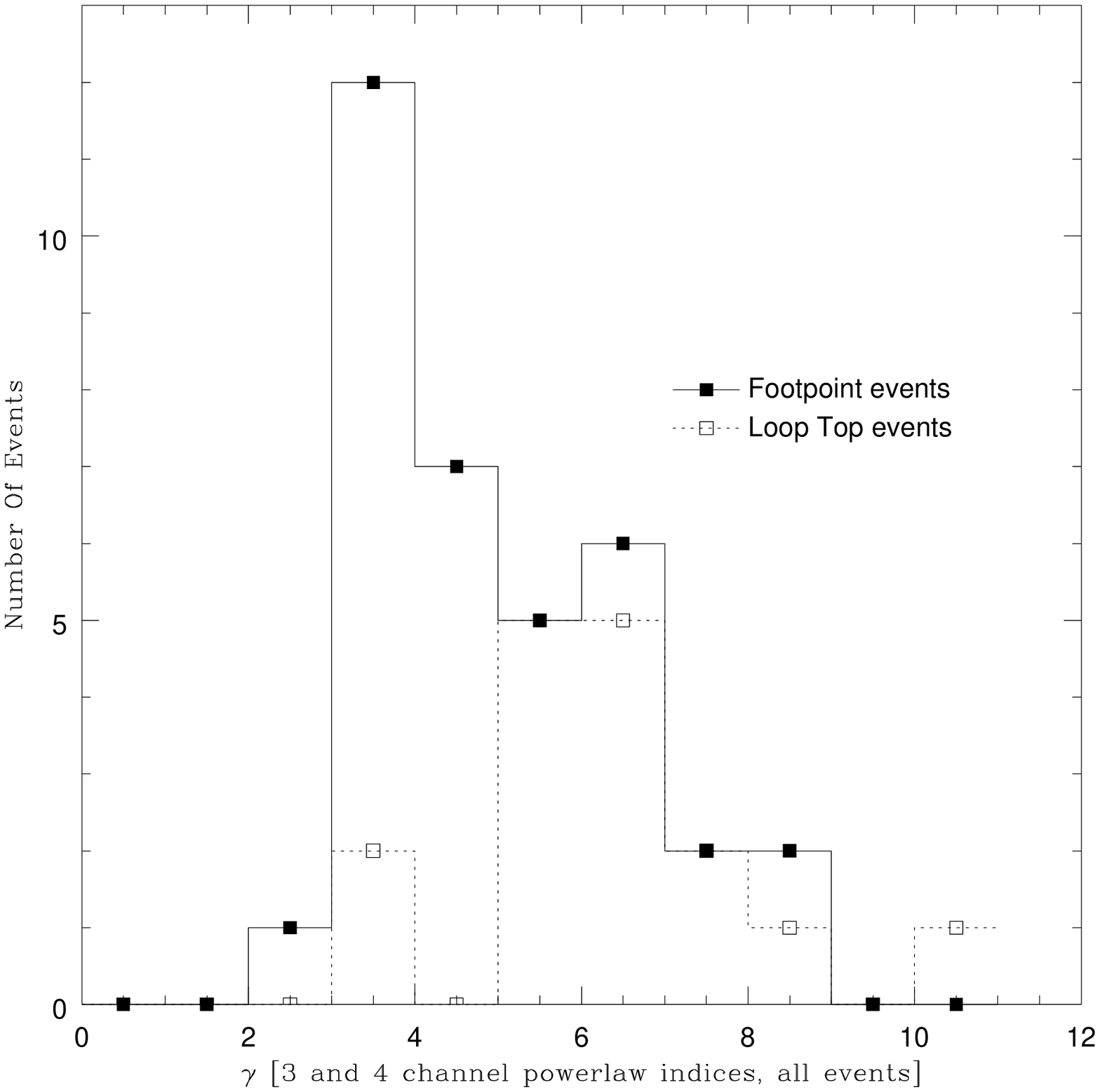,height=2.9in}
\epsfig{file=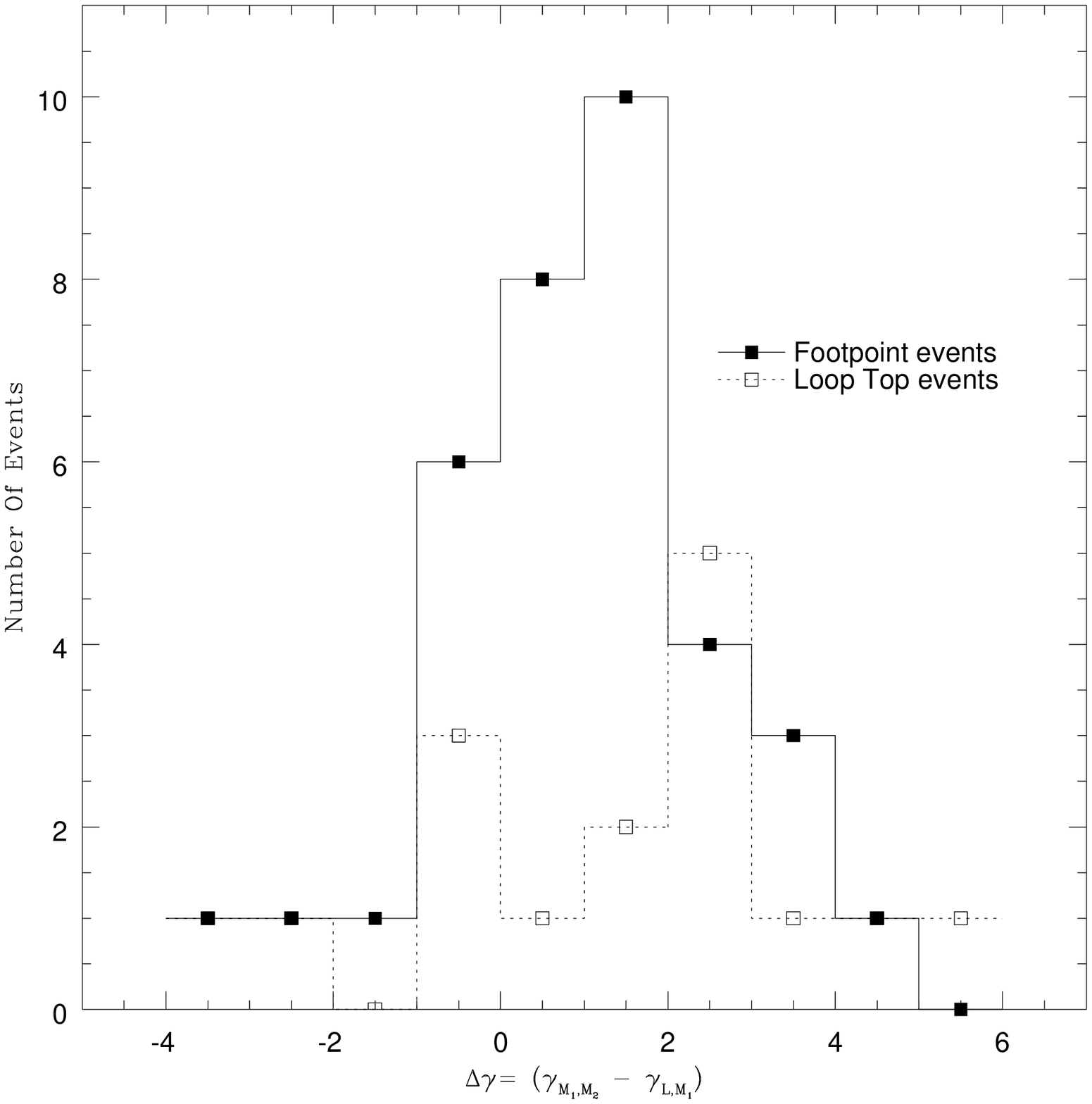,height=2.9in}
\caption{Distribution of the overall (three and four channel) index $\gamma$ 
({\bf left panel}), and the distribution of 
$\Delta\gamma=\gamma_{M1,M2}-\gamma_{L,M1}$ for all 
18 events ({\bf right panel}):
The solid histograms and filled points represent the FP sources and the 
dotted histogram and open points represent the loop 
top sources.}
\label{diffindex}
\end{figure}

Figure \ref{3chindex} shows the distribution of the spectral indices for FP 
(solid histograms) and LT (dashed histograms) sources from the simple power law  
fits, 
for 
events with no H band  images (left panel) and  events with H band  images 
(right 
panel).
Several features are readily apparent. First, the main difference between the 
two 
samples is absence of significant number of high spectral index ($\gamma \gtrsim  
6$) events in the sample with H band images (6 vs 16 in the sample with no H 
band images) and a higher average spectral index, ${\bar \gamma}_{4ch}=4.65\pm 
1.2$ vs ${\bar \gamma}_{3ch}=5.73\pm 1.5$. This is as expected. The HXT 
detects 
roughly equal counts in four channels from a flare with index $\gamma= 4$. So 
that for $\gamma>6$ the counts in the H band will fall below that of the M2 band 
by a factor of 4. This combined with our chosen threshold of 10 cts s$^{-1}$ 
SC$^{-1}$for the 
M2 band 
indicate 
that such flares in our sample would have an H band count rate of about 2 to 3 
cts s$^{-1}$ SC$^{-1}$ (which is at one or two $\sigma$ level) and will not 
yield 
a discernible image.  (A background count rate of 2 cts s$^{-1}$ SC$^{-1}$ for 
the H band 
implies a $\sigma \sim 1$ or 2 cts s$^{-1}$ SC$^{-1}$.) Thus, as stated above, 
this 
difference is purely due to the selection bias. In this 
connection we should note that  
spectral indices larger than 6 or 7 based on minimum $\chi^2$ may not be very 
reliable. Moreover, such sources may have a significant thermal contribution 
or be a super-hot thermal source like the flare of 920206. If we ignore such 
flares (say those with $\gamma \geq 7.5$), then the overall distribution of the 
combined sample, shown on the left 
panel of Figure \ref{diffindex}, is very similar to each other (${\bar 
\gamma}_{4ch}=4.65\pm 1.2$ vs ${\bar \gamma}_{3ch}=5.03\pm 1.3$) and to
previous determinations of this distribution (see McTiernan 
\& Petrosian 1991 and references cited there; see also \S 5).
Second, despite a sizable population of flares with $\gamma \simeq$ 3.0 to 4.0 
there exists a sharp cutoff at around $\gamma \simeq 3.0$ for both 
distributions. This must be intrinsic to the sources because there is no obvious 
selection bias that can produce it. 
 
Finally, for both  (and therefore the combined) samples, the LT sources tend to 
have larger 
indices than 
the 
FP sources. This is evident in the above figures, which show a clear 
shift  of the histogram of the LT sources relative to that of FPs (spectral 
index shift of about 1) and from following average indices: For 3 channel, 4 
channel and all sources respectively we get (${\bar \gamma}_{LT}=6.5\pm 1.6, 
{\bar \gamma}_{FP}=5.3\pm 1.7$), (${\bar\gamma}_{LT}=5.5\pm 1.3, 
{\bar\gamma}_{FP}=4.3\pm 1.0$), and (${\bar\gamma}_{LT}=6.2\pm 1.5, 
{\bar\gamma}_{FP}=4.9\pm 1.5$). 

Another important feature of the above results is the distribution of the 
spectral breaks. This effect is demonstrated by the right panel of Figure 
\ref{diffindex}, where we show 
the distribution of $\Delta\gamma=\gamma_{M1,M2}-\gamma_{L,M1}$ 
for all 18 events.
The most noticeable trend in this figure is that, although there are both 
positive and negative values for $\Delta\gamma$, both 
the 
FP and LT distributions are weighted toward the positive side. The median values 
of $\Delta\gamma$ is about +1 and +2 for FP and LT sources, respectively. If we 
eliminate sources with $\gamma_{L,M1}>6$, which may suffer from thermal 
contamination, the median values remain essentially unchanged but all sources 
with $\Delta\gamma\lesssim -0.5$ are eliminated. Considering the approximate 
nature of these values, the data are consistent with all thermally 
uncontaminated 
sources showing steepenings, i.e. $\Delta\gamma>0$.   
Higher 
resolution data, extended to higher energies is required to clarify this 
picture.

As in the case of the flux ratios, the precision of our analysis is greatly 
limited by our small sample size and short spectral range coverage.
Nonetheless, the above distributions provide us with initial data points that 
can 
be used to demonstrate how one can set  constraints on the model parameters. As 
evident from the theoretical 
results presented in Figs. 3, 5 and 9 of {\bf PD}, spectral indices $\gamma$ 
greater 
than 3 imply a short escape time in comparison to the acceleration time. This is 
similar to what we inferred from the distribution of the flux ratios and
may favor the smaller ratio of pitch angle to momentum diffusion coefficients 
advocated in Pryadko \& Petrosian (1997, 1998, 1999). The larger value of the 
spectral indices of LT sources also tells us  about the acceleration 
process. This difference in the spectral indices of FP and LT sources is 
related, among other things,  to the energy dependence of the pitch angle 
diffusion coefficient and/or the escape time. As shown by equation (15) in {\bf 
PD}, 
$\gamma_{LT}-\gamma_{FP}=2-s/2$, where $s$ is an index describing the energy 
dependence of the escape time (see Eq. [5] in {\bf PD}). Thus, the average value 
of about 1 for this difference seen in Figures \ref{3chindex} and 
\ref{diffindex} implies a value of 
$s=2$. Inspection of the above tables shows that the  spectral difference 
for specific FP and LT  sources associated with each other varies from 0 to 2 
indicating
a wide range of  0 to +4 for $s$ in this simple parametrization. 
Clearly, these numbers cannot be taken too seriously but 
demonstrate how model parameters can be constrained using  more refined data 
such as that expected from HESSI. 

\section{CONCLUSIONS}

We have used The YOHKOH HXT Image Catalogue (Sato et al. 1988) to determine the 
frequency of occurrence of LT hard X-ray impulsive emission in solar 
flares, and to compare the flux and spectral characteristics of these sources 
with those of the commonly observed FP sources. We have used Masuda's 
(1994) selection criteria and the YOHKOH spectral and spatial analyses software 
packages. In a few cases we have also used the Alexander \& Metcalf (1997) 
``pixon" 
method of image reconstruction. We determine the relative fluxes of the LT 
and FP sources and obtain rough measures for some of  the spectral 
characteristics (e.g. Power-law indices) 
using all possible ways of analysis that a 4 or 3 channel data set will allow.  
Analysis of these results lead us to the following conclusions:

1. The LT hard X-ray emission seems to be a common characteristic of 
the impulsive phase of 
solar flares, appearing in some form in 15 of the 18 selected flares. The 
absence 
of LT emission in the remaining cases is most likely due to the finite 
dynamic range of the imaging technique. 

2. The ratio of the summed FP to LT fluxes, lies in the range  $10 \gtrsim 
{\cal R}=FPs/LT \gtrsim 1$ and has a relatively flat distribution in this range. 
The lower limit is intrinsic to the process but the upper 
limit is most likely due to the finite (one decade)  dynamic range of the 
imaging technique. Because of this it is difficult to know the true 
distribution of this ratio.

3. The overall distribution of the power-law spectral indices rises rapidly 
above $\gamma=2$, peaks 
around 4 and declines gradually thereafter. This is similar to previous 
determinations of this distributions from HXRBS on board the {\it Solar Maximum 
Mission} (McTiernan \& Petrosian 1991), but contains a few more steep spectra, 
specially for LT sources, which could be due to thermal contamination and 
lower energy range of the  HXT compared to HXRBS. This agreement is encouraging 
and 
indicates the reliability of our spectral determination.

4. The spectra tend to steepen at higher energies (spectral index $\gamma$ 
increases by 1 to 
2), especially for sources with $\gamma <6$, for which the 
thermal 
contribution should be the lowest. This is the opposite of what is observed at 
higher energies, where spectra tend to flatten above 100's of keV (McTiernan \& 
Petrosian 1991). The directivity  of the X-ray emission and the albedo effect 
for the limb flares 
under 
consideration could also play some role, especially for the FP 
source (in the model discussed by {\bf PD} the LT source should emit 
isotropically). However, this is expected to be a small effect at low energies 
($<100$ keV) under consideration here.

5.  The spectral index $\gamma$ of LT sources is larger ({\it i.e.}  spectra are
steeper) than that of the FP sources on the average by one 
(${\bar\gamma}_{LT}=6.2\pm 1.5, {\bar\gamma}_{FP}=4.9\pm 1.5$).  The differences 
between
directivity and the albedo effects mentioned above could be a partial reason for
such a behavior, but the physics of the acceleration process must certainly play
a role here.

6. We have also described how  the above results can be used to constrain the 
model parameters especially those related to the acceleration process. For 
example, for models with acceleration at the LT source (see, e.g. Park, 
Petrosian \& Schwartz 1997 and {\bf PD}), the observed ranges of the flux ratios 
and spectral indices indicate a  rapid escape of the accelerated electron  
relative to the acceleration timescale, which is related to the momentum 
diffusion coefficient in the acceleration process. Moreover, the difference 
between 
spectral indices of LT and FP sources can constrain the energy 
dependence of the escape time which is related to the pitch angle diffusion rate 
in the acceleration site.

This demonstrates that similar studies of these characteristics of the flares 
can yield important 
information about the 
genesis and evolution of solar flares, and we eagerly anticipate the increased 
spectral, temporal and spatial resolution possible with the instruments of the 
upcoming HESSI satellite.

Finally, we note that solar flares occur in many different morphologies, the 
most common being a simple flaring loop with one LT and two FP 
sources.
However, as we discussed in \S 3.2, interacting loop models and even more 
complicated structures are frequently observed.

We would like to thank an anonymous referee for a careful reading of the 
original manuscript and for numerous helpful comments and suggestions that 
improved the paper considerably. This work is supported in parts by NASA grants 
NAG-5-7144-0002 and  
NAG5-8600-0001 and by a fellowship to TQD from 
Stanford's Undergraduate Research Opportunities.

\newpage

\end{document}